\documentclass[submitting]{nst}

\usepackage{subfigure,dcolumn}
\usepackage{epstopdf}
\usepackage{mhchem}
\usepackage{upgreek}

\begin{document}

\title{How transverse momentum conservation breaks azimuthal correlation factorization}
\thanks{Supported by the National Natural Science Foundation of China under Grants  No. 12325507, No. 12547102, and No. 12147101, and the National Key Research and Development Program of China under Grant No. 2022YFA1604900 (J.P. and G.M.), the Ministry of Science and Higher Education (PL), and the National Science Centre (PL), Grant No. 2023/51/B/ST2/01625 (A.B.)}

\author{Jia-Lin Pei}
\affiliation{Key Laboratory of Nuclear Physics and Ion-beam Application~(MOE), Institute of Modern Physics, Fudan University, Shanghai $200433$, China}
\affiliation{Shanghai Research Center for Theoretical Nuclear Physics, NSFC and Fudan University, Shanghai $200438$, China}

\author{Guo-Liang Ma }
\email{glma@fudan.edu.cn}
\affiliation{Key Laboratory of Nuclear Physics and Ion-beam Application~(MOE), Institute of Modern Physics, Fudan University, Shanghai $200433$, China}
\affiliation{Shanghai Research Center for Theoretical Nuclear Physics, NSFC and Fudan University, Shanghai $200438$, China}

\author{Adam Bzdak}
\email{bzdak@fis.agh.edu.pl}
\affiliation{AGH University of Krakow,\\
Faculty of Physics and Applied Computer Science,
30-059 Krak\'ow, Poland}

\begin{abstract}
The breakdown of azimuthal two-particle correlation factorization, quantified by the ratios $r_2$ and $r_3$, serves as a sensitive probe of transverse-momentum-dependent flow fluctuations. While hydrodynamic models predict $r_3 \leq 1$, experimental data from CMS in p-Pb collisions exhibit $r_3 > 1$, presenting a clear puzzle. We show that transverse momentum conservation (TMC) is the key mechanism dictating this factorization breakdown in small systems. We systematically calculate the effect of TMC as a function of the momentum difference between particles across various multiplicity and momentum ranges. Our results are in quantitative agreement with CMS p-Pb data for both $r_2$ and $r_3$. A central finding is a sign rule: under TMC, the deviation $r_n - 1$ follows $\left ( - 1 \right )^{n+1}  $, being negative for even and positive for odd harmonic orders $n$. This work quantifies a dominant non-flow contribution to the factorization ratios $r_n$ in small systems, and demonstrates that this kinematic effect must be subtracted or accounted for before $r_n$ can be used as a reliable probe of genuine flow fluctuations.
\end{abstract}


\maketitle

\section{Introduction}
The quark-gluon plasma (QGP), a deconfined state of quarks and gluons governed by Quantum Chromodynamics (QCD) under conditions of extreme temperature and density, is the primary object of study in ultra-relativistic heavy-ion collisions at the BNL Relativistic Heavy Ion Collider (RHIC) and the CERN Large Hadron Collider (LHC)~\cite{LHC1,nstygma,nst4,LHC2,LHC3,xpduan,nstQCD,nstxpduan}. Among the most powerful signatures of this state is the azimuthal anisotropy of particle emission in the transverse plane~\cite{obser1,obser2,obser3,obser4}. The initial geometric asymmetries and energy-density inhomogeneities are mapped, via the collective evolution of the QGP, into momentum anisotropies imprinted on final-state particles, a phenomenon referred to as collective flow~\cite{transf1,transf2}. This evolution is successfully described by both relativistic hydrodynamic models~\cite{model1,transf2} and transport models~\cite{Linampt}.  Motivated by such models, this anisotropy is quantified by a Fourier expansion of the azimuthal angle ($\phi$) distribution of emitted particles \cite{expan1,expan2,nst2}, 
\begin{equation}
\frac{\mathrm{d} N}{\mathrm{d} \phi } 
\propto 
1+ 2{\sum_{n}} 
v_{n}\cos[n(\phi -\Psi _{n} )] .
\label{eq1}
\end{equation}
Here, the Fourier coefficient $v_{n}$ and the event plane angle  $\Psi _{n}$ represent the magnitude and orientation, respectively, of the $n^{th}$-order harmonic flow vector $ \vec{V}_{n}= v_{n}e^{in\Psi _{n} }$. Typically, the most significant flow coefficient corresponds to the elliptic flow $v_2$, resulting from the initial almond-shaped overlap region in noncentral collisions. However, positional fluctuations of both the colliding nuclei and the nucleons within them can lead to more complex geometric configurations or higher-order eccentricities $\varepsilon _{n}$ of the initial geometry, thereby resulting in non-vanishing flow coefficients $v_n$ for $n > 2$,
such as the triangular flow $v_3$ \cite{3flow1,3flow2,3flow3,nstfluc}. The event-plane angle $\Psi_n$ is determined by the $n$th-order participant-plane angle of the initial nucleon overlap region, which fluctuates event-by-event due to the varying spatial distribution of participating nucleons~\cite{3flow1}.
The azimuthal anisotropies serve as powerful probes for constraining the initial state of heavy-ion collisions and quantifying certain transport properties of the QGP, e.g., the ratio of shear viscosity to entropy density, $\eta /s$ \cite{shear1,shear2,shear3,shear4}. Comparisons with theoretical models have revealed that the QGP behaves as a nearly perfect fluid, with a shear viscosity to entropy density ratio $\eta/s$ close to the quantum lower bound of $1/(4\pi)$. This low value signals extremely strong interactions among its constituents inside the QGP~\cite{fluid1,fluid2,nstqgp}.

A standard experimental method determines the single-particle flow harmonic $v_n$ by measuring two-particle azimuthal correlations, avoiding the uncertainties associated with event-plane estimation~\cite{2particlecorr}.
The two-particle distribution in relative azimuthal angle $\Delta \phi= \phi^{a} - \phi ^{b}$ can be expressed as a Fourier series ~\cite{expan1,expan2}, 
\begin{equation}
    \frac{\mathrm{d} N^{\text{pair}} }{\mathrm{d} \Delta \phi } 
\propto 
1+ 2{\sum_{n} } 
V_{n\Delta }\cos [n (\phi^{a} - \phi ^{b})].
\end{equation}
 If collective flow serves as the primary source of final-state particle correlations, the factorization of the two-particle correlation $V_{n\Delta }$ can be characterized as,
\begin{equation}
    V_{n\Delta }\left ( p_{T}^{a},p_{T}^{b}  \right ) =
v_{n} \left ( p_{T}^{a}  \right )
v_{n} \left ( p_{T}^{b}  \right ),
\label{eq3}
\end{equation}
where $v_{n} \left ( p_{T}^{a}  \right )$ and $v_{n} \left ( p_{T}^{b}  \right )$ represent the single-particle flow coefficients for a pair of particles with transverse momenta $ p_{T}^{a} $ and $ p_{T}^{b} $, respectively. Thus, the $ p_{T}$-dependent behavior of the single-particle $ v_{n} $ can be derived by fixing one particle within a relatively wide $ p_{T}$ interval, adjusting only the $ p_{T}$ of the other particle, and measuring $V_{n\Delta }$. 
The validity of Eq.~\eqref{eq3} requires the condition $\Psi _{n}\left ( p_{T}^{a}  \right )$=$\Psi _{n}\left ( p_{T}^{b}  \right )$=$\Psi _{n}$, whereby the common event-plane angle cancels in two-particle correlations. However, research findings have shown that in the most central Pb-Pb collisions, the factorization assumption of Eq.~\eqref{eq3} does not hold, and the degree of deviation reaches up to $20\%$ \cite{deviation}.
This is primarily attributed to the presence of local fluctuations in the energy density within the overlapping region of two nuclei at the initial stage of collisions, a feature that 
hydrodynamic models highlight. The $p_T$-differential impact of these fluctuations on particle emission causes the event plane to fluctuate across $p_T$ ranges~\cite{brokereason1,brokereason2}, thereby modifying Eq.~\eqref{eq3} as follows,
\begin{equation}
    V_{n\Delta }\left ( p_{T}^{a},p_{T}^{b}  \right ) =
v_{n} \left ( p_{T}^{a}  \right )
v_{n} \left ( p_{T}^{b}  \right ) \cos [n (\Psi(p_{T}^{a}) - \Psi(p_{T}^{b})].
\label{eq3p5}
\end{equation}
Thus, Eq.~\eqref{eq3p5} implies a $p_T$-dependent factorization breaking of Eq.~\eqref{eq3}, with the factorization ratio $r_n$ (defined in Sec. \ref{sec2}) of  $\cos[n(\Psi(p_T^a) - \Psi(p_T^b))]$. This provides a unique probe into the fluctuations in the initial state of heavy-ion collisions along the radial direction. 
 
In contrast to large A+A systems, the origin of azimuthal particle correlations in small systems ($p$+$p$ and $p$+A) remains an open and vital question~\cite{guoshuang,hswang}. The theoretical landscape for explaining azimuthal particle correlations in small systems comprises competing paradigms, including the color glass condensation (CGC) models~\cite{cgc1,cgc2,cgc3}, hydrodynamic models~\cite{hydro1,hydro2,hydro3}, and transport models like AMPT~\cite{ampt1,ampt2,nstampt}. The observed factorization breakdown also serves as a critical test of the origin(s) of azimuthal correlations in small systems, because if such effects arise from hydrodynamics, hydrodynamic models must account for the experimental data. The factorization ratios $r_n$ for $V_{n\Delta}$ ($n$=2 and 3), to be defined in Eq.~(\ref{flowrn}), have been measured in p-Pb collisions by the CMS collaboration~\cite{cmsdata}. For the elliptic harmonic ($V_{2\Delta}$), hydrodynamic simulations with MC-Glauber initial conditions can approximately describe the data, where the measured factorization ratio $r_2$ falls below unity at low multiplicity and high $p_T$. In contrast, the triangular harmonic ($V_{3\Delta}$) exhibits a ratio $r_3$ that exceeds unity in the same kinematic region, incompatible with hydrodynamic calculations which predict $r_3 \leq 1$. This discrepancy demonstrates that hydrodynamic models cannot simultaneously describe the factorization patterns of both $r_2$ and $r_3$ in small systems.
In addition, a systematic study of $p_T$-dependent flow vector fluctuations within the AMPT model demonstrated that the factorization breaking of $V_{2\Delta}$ is highly sensitive to changes in initial conditions, but exhibits negligible dependence on the parton cross-section and hadronic rescatterings~\cite{amptzy}. This underscores the need to explore the mechanisms of factorization breaking, as precision studies of this phenomenon in small systems will thus be helpful for understanding the origin(s) of azimuthal particle correlations in small systems~\cite{smallsys1}. 

Transverse momentum conservation (TMC), a distinct class of "non-flow" contribution, imposes an inherent constraint on final-state momentum distributions, generating significant azimuthal correlations that are particularly important in low-multiplicity small systems~\cite{tmc1,tmc2,tmc3}. Our earlier work has shown that the interplay between the TMC and collective flow can describe a range of observed correlations in small systems~\cite{grouptmc1,grouptmc2,grouptmc3,grouptmc4,grouptmc5}. In this paper, we calculate the factorization breakdown of $V_{2\Delta }$ and $V_{3\Delta } $ based on the TMC framework. By quantifying this non-flow imprint on the factorization ratios $r_2$ and $r_3$, our study provides a critical framework for disentangling genuine flow vector fluctuations and non-flow effects in experimental measurements. We compare our analytical calculations with CMS p-Pb data~\cite{cmsdata} to explore the mechanisms driving azimuthal particle correlations in small systems.

\section{factorization ratios from TMC and flow}
\label{sec2}

The observable used to study $p_T$-dependent flow vector fluctuations is the factorization ratio $r_n$, which is defined via two-particle harmonic flows and used to probe the factorization of two particles in different $p_T$ intervals.  Its definition is given by \cite{defin},
\begin{equation}
    r_{n}=\frac{V_{n\Delta }\left ( p_{T}^{a},p_{T}^{b} \right )  }
{\sqrt{V_{n\Delta }\left ( p_{T}^{a},p_{T}^{a} \right )V_{n\Delta }\left ( p_{T}^{b},p_{T}^{b} \right )} }  ,
\label{defrn}
\end{equation}
where 
\begin{equation}
   V_{n\Delta }\left ( p_{T}^{a},p_{T}^{b} \right ) =\left \langle
 e^{in\left ( \phi _{1}^{a}-\phi _{2}^{b}  \right ) }  \right \rangle   
= \left \langle \cos [n\left ( \phi _{1}^{a}-\phi _{2}^{b}  \right )] \right \rangle ,
\end{equation}
and the denominator represents the scenario where two particles originate from the same $p_T$ interval, i.e.
\begin{equation}
\begin{aligned}
    V_{n\Delta }\left ( p_{T}^{a},p_{T}^{a} \right ) = \left \langle
 e^{in\left ( \phi _{1}^{a}-\phi _{2}^{a}  \right ) }  \right \rangle   
&= \left \langle \cos [n\left ( \phi _{1}^{a}-\phi _{2}^{a}  \right )] \right \rangle  \\ 
&= v_{n}\left ( p_{T}^{a}  \right )^{2},
\end{aligned}
\end{equation}
\begin{equation}
\begin{aligned}
    V_{n\Delta }\left ( p_{T}^{b},p_{T}^{b} \right ) = \left \langle
 e^{in\left ( \phi _{1}^{b}-\phi _{2}^{b}  \right ) }  \right \rangle   
&= \left \langle \cos [n\left ( \phi _{1}^{b}-\phi _{2}^{b}  \right )] \right \rangle  \\ 
&= v_{n}\left ( p_{T}^{b}  \right )^{2}.
\end{aligned}
\end{equation}
If collective flow dominates and non-flow effects are negligible, Eq.~\eqref{defrn} will be equivalent to the following formula,
\begin{equation}
    r_{n}=\frac{ v_{n}\left ( p_{T}^{a}  \right )
v_{n}\left ( p_{T}^{b}  \right ) \cos \left [ n(\Psi _{n}(p_{T}^{a})-\Psi _{n}(p_{T}^{b}) ) \right ] }
{\sqrt{v_{n}\left ( p_{T}^{a}  \right )^{2}
v_{n}\left ( p_{T}^{b}  \right )^{2} }
 } .
 \label{flowrn}
\end{equation}
Consequently, the Cauchy-Schwarz inequality gives $r_n \le 1$ for the factorization ratio defined in Eq.~\eqref{flowrn}~\cite{cauchy}.

\subsection{TMC calculations}
\label{TMC calculations}
Consider a collision producing a total of 
$N$ particles, with their individual transverse momenta being $\vec{p}_{1},\vec{p}_{2},\dots , \vec{p}_{N} $ (For the sake of brevity, we use $p$ to represent $p_T$ in the subsequent calculations). Under the constraint of transverse momentum conservation, the transverse momentum distribution (normalized to unity) $f_N$ of these $N$ particles can be described as \cite{grouptmc2,tmcmethod2,tmcmethod3,tmcmethod4}
\begin{widetext}
\begin{equation}
 \label{fN}
\begin{aligned}
 f_{N}(\vec{p}_{1},\vec{p}_{2},\dots , \vec{p}_{N}) =
\frac{\delta ^{2}(\vec{p}_{1}+\vec{p}_{2}+\dots +\vec{p}_{N}) \prod_{i=1}^{N} f(\vec{p}_{i})}
{\int_{F}\delta ^{2}(\vec{p}_{1}+\vec{p}_{2}+\dots +\vec{p}_{N}) \prod_{i=1}^{N} \left [ f(\vec{p}_{i})d^{2}\vec{p_i} \right ] 
 } ,  
\end{aligned}
\end{equation}
where the $\delta$ function constitutes the mathematical expression of TMC, $f(\vec{p}_{i})$ is the normalized single-particle distribution, and it is assumed here that all particles follow the same transverse momentum distribution; $F$ denotes the full phase space. To calculate two-particle azimuthal correlation function such as $\left \langle \cos \left [ n(\phi _{1}-\phi_{2} ) \right ]  \right \rangle $, it is necessary to know the two-particle distribution function $f_2$, which can be obtained by integrating out the momenta of the remaining $M=N-2$ particles in Eq.~\eqref{fN},
\begin{equation}
 \label{f2}
\begin{aligned}
 f_{2}(\vec{p}_{1},\vec{p}_{2}) =f(\vec{p}_{1})f(\vec{p}_{2})
\frac{\int_{F}\delta ^{2}(\vec{p}_{1}+\vec{p}_{2}+\dots +\vec{p}_{N}) \prod_{i=3}^{N}\left [ f(\vec{p}_{i})d^{2}\vec{p_i} \right ] }
{\int_{F}\delta ^{2}(\vec{p}_{1}+\vec{p}_{2}+\dots +\vec{p}_{N}) \prod_{i=1}^{N} \left [ f(\vec{p}_{i})d^{2}\vec{p_i}\right ] 
 }.
\end{aligned}
\end{equation}
To evaluate the integral in Eq.~\eqref{f2}, we define the sum of the transverse momenta of $M$ particles as $\vec{K}\equiv  {\textstyle \sum_{i=1}^{M}}\vec{p}_{i}   $. According to the central limit theorem, $\vec{K}$ satisfies a Gaussian distribution, i.e.,
\begin{equation}
 \label{CLT}
\begin{aligned}
G_{M}(\vec{K})\equiv \displaystyle\int_{F}   \delta ^{2}\left ( -\vec{K}+\sum_{i=1}^{M}\vec{p}_{i} \right )
\displaystyle \prod_{i=1}^{M}\left (f\left ( \vec{p}_{i} \right )d^{2}\vec{p}_{i}  \right ) 
= \frac{1}{\pi\sigma ^{2} }\text{exp}\bigg( -\frac{\mid\vec{K}\mid ^{2} }{\sigma ^{2}} \bigg),
\end{aligned}
\end{equation}
where the mean and variance of $\vec{K}$ are as follows:
\begin{equation}
\begin{aligned}
\left \langle \vec{K} \right \rangle_{F} =\left \langle \sum_{i=1}^{M}\vec{p}_{i} \right \rangle_{F}
= \sum_{i=1}^{M}\left \langle \vec{p}_{i} \right \rangle_{F} &=0 ,\hspace{0.5cm}
\sigma ^{2}=M\left ( \left \langle \left | \vec{p} \right |^{2}   \right \rangle_{F}-\left \langle \vec{p} \right \rangle^{2}_{F} \right ) 
= M\left \langle \left | \vec{p} \right |^{2}   \right \rangle_{F}, 
\end{aligned}
\end{equation}
and
\begin{equation}
\begin{aligned}
 \left \langle \left | \vec{p} \right | ^{2}\right \rangle_{F} =
\frac{\int_{F} \left | \vec{p} \right |^{2}f(\vec{p})d^{2}\vec{p}}{\int_{F} f(\vec{p})d^{2}\vec{p}}.
\end{aligned}
\end{equation}
Therefore, in accordance with Eq.~\eqref{CLT}, Eq.~\eqref{f2} can be further described as follows:
\begin{equation}
    f_{2}(\vec{p}_{1},\vec{p}_{2})= f(\vec{p}_{1})f(\vec{p}_{2})
\frac{G_{N-2}\left (- \vec{p}_{1}- \vec{p}_{2} \right )   }
{G_{N}\displaystyle\left ( 0 \right )}=
f(\vec{p}_{1})f(\vec{p}_{2})
\frac{N}{N-2} \text{exp}\left ( -\frac{ \left ( \vec{p}_{1}+ \vec{p}_{2}  \right )^{2} }
{(N-2)\left \langle \left | \vec{p} \right |^{2}  \right \rangle_{F} }  \right ).
\label{f2s}
\end{equation}
It can be seen from Eq.~\eqref{f2s} that the two-particle density distribution $f_2$ depends on the sum of the transverse momentum vectors, and thus reaches its maximum value in the back-to-back configuration. In addition, we also need to specify the form of the single-particle distribution, which is generally determined as follows:
\begin{equation}
    f\left ( \vec{p} \right )=\frac{g\left ( p \right ) }{2\pi}\left ( 1+2v_{2} \left (  p\right ) 
\cos \left [ 2\left ( \phi -\Psi _{2}(p)  \right )  \right ]   \right ).
\label{singlef}
\end{equation}
By combining Eq.~\eqref{singlef}, Eq.~\eqref{f2s} can be fully expressed as:
\begin{equation}
 \begin{split}
        f_{2}(\vec{p}_{1},\vec{p}_{2})=&\frac{g\left ( p_{1}  \right ) }{2\pi}\frac{g\left ( p_{2}  \right ) }{2\pi}\left ( 1+2v_{2} \left (  p_{1} \right ) 
\cos \left [ 2\left ( \phi_{1}  -\Psi _{2}(p_{1} )  \right )  \right ]   \right )
\left ( 1+2v_{2} \left (  p_{2} \right ) 
\cos \left [ 2\left ( \phi_{2}  -\Psi _{2}(p_{2} )  \right )  \right ]   \right ) \\
& \hspace{0.15cm} \times  \frac{N}{N-2} \mathrm{exp}\left ( - \frac{\left ( p_{1,x}+p_{2,x} \right )^{2} }
{2\left (  N-2\right )\left \langle p_{x}^{2}  \right  \rangle _{F} }
  - \frac{\left ( p_{1,y} +p_{2,y} \right )^{2}  }
{2\left (  N-2\right )\left \langle p_{y}^{2} \right \rangle _{F} }\right ),
\end{split}
\label{f2s1}
\end{equation}
where
\begin{equation}
\begin{aligned}
p_{x} = p\cos \left (  \phi\right ),
\hspace{0.5cm}p_{y} = p\sin \left (  \phi\right ),
\end{aligned}
\end{equation}
\begin{equation}
\left \langle p_{x}^{2}   \right \rangle_{F}  = \frac{1}{2} \left \langle p^{2}  \right \rangle _{F}\left ( 1+  v_{2F}   \right ),
\hspace{0.5cm}\left \langle p_{y}^{2}   \right \rangle_{F}  = \frac{1}{2} \left \langle p^{2}  \right \rangle _{F}\left ( 1-  v_{2F}   \right ),
\end{equation}
\begin{equation}
v_{2F}=   \frac{\int_{F} v_{2}\left ( p \right )g(p)\cos \left [ 2\Psi _{2}\left ( p \right )   \right ]p^{2}d^{2}p    }{\int_{F} g(p)p^{2}d^{2}p}.
\end{equation}
Based on Eq.~\eqref{f2s1}, we can calculate various types of two-particle azimuthal correlations under a given transverse momentum, such as $r_2$ and $r_3$ in Eq.~\eqref{defrn},
\begin{equation}
    r_{2}=\frac{\left \langle \cos [2\left ( \phi _{1}^a-\phi _{2}^b \right )] \right \rangle }
{\sqrt{\left \langle \cos [2\left ( \phi _{1}^a-\phi _{2}^a\right )] \right \rangle
\left \langle \cos [2\left ( \phi _{1}^b-\phi _{2}^b \right )] \right \rangle } }  ,
\hspace{0.5cm}
r_{3}=\frac{\left \langle \cos [3\left ( \phi _{1}^a-\phi _{2}^b \right )] \right \rangle }
{\sqrt{\left \langle \cos [3\left ( \phi _{1}^a-\phi _{2}^a\right )] \right \rangle
\left \langle \cos [3\left ( \phi _{1}^b-\phi _{2}^b \right )] \right \rangle } }  .
\label{r2r3}
\end{equation}
Taking the numerator of $r_2$ in Eq.~\eqref{r2r3} as an example, its calculation process and result are as follows:
\begin{equation}
\begin{aligned}
    \left \langle \cos [2\left ( \phi _{1}^a-\phi _{2}^b \right )] \right \rangle\mid p_{a},p_{b} &=
\frac{\int_{0}^{2\pi}\cos [2\left ( \phi _{1}^a-\phi _{2}^b \right )]
f_{2} \left ( p_{a},\phi _{1}^a,p_{b},\phi _{2}^b  \right )d\phi _{1}^a d\phi _{2}^b}
{\int_{0}^{2\pi}f_{2} \left ( p_{a},\phi _{1}^a,p_{b},\phi _{2}^b  \right )d\phi _{1}^a d\phi _{2}^b}  \\
&\approx A_{0}+A_{1}Y_{A}+\frac{1}{2} A_{2}Y_{A}^{2},
\label{r2n}
\end{aligned}
\end{equation}
where
\begin{equation}
Y _{A}=-\frac{1 }{(N-2) \left \langle p^{2}  \right \rangle _{F}},
\end{equation}
and
\begin{equation}
\label{A}
\begin{aligned}
  A_0&=v_2(p_a) v_2(p_b) \cos\left[2 \Psi_2(p_a) - 2 \Psi_2(p_b)\right],  \\
    A_1&=-\frac{1}{2} p_b^2 v_{2F} v_2(p_a) \cos\left[2 \Psi_2(p_a)\right] + p_a^2 v_2(p_a) v_2(p_b) \cos\left[2 \Psi_2(p_a) - 2 \Psi_2(p_b)\right] \\
    &\hspace{0.45cm}+ p_b^2 v_2(p_a) v_2(p_b) \cos\left[2 \Psi_2(p_a) - 2 \Psi_2(p_b)\right]- \frac{1}{2} p_a^2 v_{2F} v_2(p_b) \cos\left[2 \Psi_2(p_b)\right], \\  
     A_2&=p_a^2 p_b^2 + \frac{1}{2} p_a^2 p_b^2 v_{2F}^2 
- 3 p_a^2 p_b^2 v_{2F} v_2(p_a) \cos\left[2 \Psi_2(p_a)\right]
 - p_b^4 v_{2F} v_2(p_a) \cos\left[2 \Psi_2(p_a)\right] \\ 
&\hspace{0.45cm} + p_a^4 v_2(p_a) v_2(p_b) \cos\left[2 \Psi_2(p_a) - 2 \Psi_2(p_b)\right]
 + 4 p_a^2 p_b^2 v_2(p_a) v_2(p_b) \cos\left[2 \Psi_2(p_a) - 2 \Psi_2(p_b)\right] 
 \\
&\hspace{0.45cm}+ p_b^4 v_2(p_a) v_2(p_b) \cos\left[2 \Psi_2(p_a) - 2 \Psi_2(p_b)\right] + \frac{1}{2} p_a^4 v_{2F}^2 v_2(p_a) v_2(p_b) \cos\left[2 \Psi_2(p_a) 
- 2 \Psi_2(p_b)\right] \\
&\hspace{0.45cm}+ 2 p_a^2 p_b^2 v_{2F}^2 v_2(p_a) v_2(p_b) 
\cos\left[2 \Psi_2(p_a) - 2 \Psi_2(p_b)\right]+ \frac{1}{2} p_b^4 v_{2F}^2 v_2(p_a) v_2(p_b) \cos\left[2 \Psi_2(p_a)- 2 \Psi_2(p_b)\right] \\
&\hspace{0.45cm}+\frac{1}{4} p_a^4 v_{2F}^2 v_2(p_a) v_2(p_b) \cos\left[2 \Psi_2(p_a) 
+ 2 \Psi_2(p_b)\right] + \frac{1}{4} p_b^4 v_{2F}^2 v_2(p_a) v_2(p_b) \cos\left[2 \Psi_2(p_a) 
+ 2 \Psi_2(p_b)\right] \\
&\hspace{0.45cm}- p_a^4 v_{2F} v_2(p_b) \cos\left[2 \Psi_2(p_b)\right]- 3 p_a^2 p_b^2 v_{2F} v_2(p_b) \cos\left[2 \Psi_2(p_b)\right].
\end{aligned}
\end{equation}
The above results are based on the expansion of the exponential term in Eq.~\eqref{f2s1} to the second order, i.e., $\text{exp} \left ( -X \right ) \approx 1-X+\frac{X^{2}}{2} $, since the first non-vanishing pure TMC term, i.e., the first term of $A_2$ in Eq.~\eqref{A}, appears at $\frac{X^{2}}{2}$. For the denominator of Eq.~\eqref{r2n}, we only take the first term of its expansion, so the result is $\left ( 2\pi \right ) ^{2} $. For the complex results in Eq.~\eqref{A}, we categorize it into three distinct classes according to their physical origin: 

\vspace{\baselineskip}
\noindent Type I (pure flow): 
Terms that represent collective flow effects, which depend only on $v_{n}$ and $\Psi _{n}$, i.e. $A_0$.

\vspace{\baselineskip}
\noindent
Type II (pure TMC): Terms that arise solely from the transverse momentum conservation constraint, which depend only on $N$ and $p$, i.e. the first term of $A_2$.

\vspace{\baselineskip}
\noindent
Type III (interplay): Terms that describe the interplay between pure flow and pure TMC, which depend on all of $v_{n}$, $\Psi _{n}$, $N$, and $p$, i.e., $A_1$ and $A_2$ except the first term. \\
 To simplify the analysis, we retain only the leading term of $A_2$ in Eq.~\eqref{A}. This is justified because $A_2$ is suppressed by a factor of $1/N^2$, while all non-leading terms carry an additional suppression factor of $v_n^2$.
Furthermore, the remaining interplay term $A_1$ is numerically negligible in the experimentally relevant regime, as demonstrated by the direct comparison in Sec. \ref{secB}. With these approximations, only the pure flow term and the pure TMC term dominate in ~\eqref{A}, leading to:
\begin{equation}
    c_{2}\left \{ 2 \right \}\!\mid _{p_{a},p_{b} }=v_2(p_a) v_2(p_b)+
  \frac{p_{a}^{2}p_{b}^{2}  }{2(N-2)^{2}  \left \langle p^{2}  \right \rangle _{F}^{2}},
\end{equation}
where the pure TMC term  $p^2_a p^2_b /\ [2(N-2)^2\langle p^2\rangle^2_F]$ is the key ingredient driving $r_2 < 1$ in the subsequent calculations.
In principle, we should also include $v_3$ in the above calculations; however, we verified that its contribution is negligible in Appendix \ref{secwithv3}.
For the denominator of $r_2$ in Eq.~\eqref{r2r3}, each factor under the square root corresponds to evaluating Eqs.~\eqref{r2n} and~\eqref{A} with both $p_a$ and $p_b$ uniformly set to $p_a$ or to $p_b$, respectively.

Similarly, for the numerator of $r_3$ in Eq.~\eqref{r2r3}, we obtain:
\begin{equation}
\begin{aligned}
    \left \langle \cos [3\left ( \phi _{1}^a-\phi _{2}^b \right )] \right \rangle\mid p_{a},p_{b} &=
\frac{\int_{0}^{2\pi}\cos [3\left ( \phi _{1}^a-\phi _{2}^b \right )]
f_{2} \left ( p_{a},\phi _{1}^a,p_{b},\phi _{2}^b  \right )d\phi _{1}^a d\phi _{2}^b }
{\int_{0}^{2\pi}f_{2} \left ( p_{a},\phi _{1}^a,p_{b},\phi _{2}^b  \right )d\phi _{1}^a d\phi _{2}^b} \\
&\approx B_{0}+B_{1}Y_{B}+\frac{1}{2} B_{2}Y_{B}^{2}+\frac{1}{6} B_{3}Y_{B}^{3},
\label{r3n}
\end{aligned}
\end{equation}
where
\begin{equation}
Y _{B}=-\frac{1 }{(N-2) \left \langle p^{2}  \right \rangle _{F}},
\end{equation}
\begin{equation}
 \begin{split}
        f_{2}(p_{a},\phi _{1}^a,p_{b},\phi _{2}^b)=&\frac{g\left ( p_{a}  \right ) }{2\pi}\frac{g\left ( p_{b}  \right ) }{2\pi}\left ( 1+2v_{2} \left (  p_{a} \right ) 
\cos \left [ 2\left ( \phi_{1}^a  -\Psi _{2}(p_{a} )  \right )  \right ] +2v_{3} \left (  p_{a} \right ) 
\cos \left [ 3\left ( \phi_{1}^a  -\Psi _{3}(p_{a} )  \right )  \right ]  \right ) \\
\times &
\left ( 1+2v_{2} \left (  p_{b} \right ) 
\cos \left [ 2\left ( \phi_{2}^b  -\Psi _{2}(p_{b} )  \right )  \right ] +2v_{3} \left (  p_{b} \right ) 
\cos \left [ 3\left ( \phi_{2}^b  -\Psi _{3}(p_{b} )  \right )  \right ]  \right ) \\
\times &\frac{N}{N-2} \mathrm{exp}\left ( - \frac{\left ( p_{a,x}+p_{b,x} \right )^{2} }
{2\left (  N-2\right )\left \langle p_{x}^{2}  \right  \rangle _{F} }
  - \frac{\left ( p_{a,y} +p_{b,y} \right )^{2}  }
{2\left (  N-2\right )\left \langle p_{y}^{2} \right \rangle _{F} }\right ),
\end{split}
\label{f2v3}
\end{equation}
and
\begin{equation}
\begin{aligned}
   B_0&=v_3(p_a) v_3(p_b) \cos\left[3 \Psi_3(p_a) - 3 \Psi_3(p_b)\right], \\
B_1&=p_a p_b v_2(p_a) v_2(p_b) \cos\left[2 \Psi_2(p_a) - 2 \Psi_2(p_b)\right] 
+ p_a^2 v_3(p_a) v_3(p_b) \cos\left[3 \Psi_3(p_a) - 3 \Psi_3(p_b)\right] \\
&\hspace{0.45cm}+ p_b^2 v_3(p_a) v_3(p_b) \cos\left[3 \Psi_3(p_a) - 3 \Psi_3(p_b)\right], \\
B_2&=-p_a p_b^3 v_{2F} v_2(p_a) \cos\left[2 \Psi_2(p_a)\right]
 + 2 p_a^3 p_b v_2(p_a) v_2(p_b) \cos\left[2 \Psi_2(p_a) - 2 \Psi_2(p_b)\right]  \\
&\hspace{0.45cm}+ 2 p_a p_b^3 v_2(p_a) v_2(p_b) \cos\left[2 \Psi_2(p_a) - 2 \Psi_2(p_b)\right] 
+ p_a^3 p_b v_{2F}^2 v_2(p_a) v_2(p_b) \cos\left[2 \Psi_2(p_a) - 2 \Psi_2(p_b)\right]  \\
&\hspace{0.45cm}+ p_a p_b^3 v_{2F}^2 v_2(p_a) v_2(p_b) \cos\left[2 \Psi_2(p_a) - 2 \Psi_2(p_b)\right]
 - p_a^3 p_b v_{2F} v_2(p_b) \cos\left[2 \Psi_2(p_b)\right]  \\
&\hspace{0.45cm} + p_a^4 v_3(p_a) v_3(p_b) \cos\left[3 \Psi_3(p_a) - 3 \Psi_3(p_b)\right] 
+ 4 p_a^2 p_b^2 v_3(p_a) v_3(p_b) \cos\left[3 \Psi_3(p_a) - 3 \Psi_3(p_b)\right] \\
&\hspace{0.45cm}+ p_b^4 v_3(p_a) v_3(p_b) \cos\left[3 \Psi_3(p_a) - 3 \Psi_3(p_b)\right] 
+ \frac{1}{2} p_a^4 v_{2F}^2 v_3(p_a) v_3(p_b) \cos\left[3 \Psi_3(p_a) - 3 \Psi_3(p_b)\right] \\
&\hspace{0.45cm}+ 2 p_a^2 p_b^2 v_{2F}^2 v_3(p_a) v_3(p_b) \cos\left[3 \Psi_3(p_a) - 3 \Psi_3(p_b)\right]
+ \frac{1}{2} p_b^4 v_{2F}^2 v_3(p_a) v_3(p_b) \cos\left[3 \Psi_3(p_a) - 3 \Psi_3(p_b)\right], \\
B_3&=p_a^3 p_b^3 + \frac{3}{2} p_a^3 p_b^3 v_{2F}^2 
- 6 p_a^3 p_b^3 v_{2F} v_2(p_a) \cos\left[2 \Psi_2(p_a)\right]- 3 p_a p_b^5 v_{2F} v_2(p_a) \cos\left[2 \Psi_2(p_a)\right] \\
&\hspace{0.45cm}- \frac{3}{2} p_a^3 p_b^3 v_{2F}^3 v_2(p_a) \cos\left[2 \Psi_2(p_a)\right]
 - \frac{3}{4} p_a p_b^5 v_{2F}^3 v_2(p_a) \cos\left[2 \Psi_2(p_a)\right] + 3 p_a^5 p_b v_2(p_a) v_2(p_b) \cos\left[2 \Psi_2(p_a) - 2 \Psi_2(p_b)\right] \\
 &\hspace{0.45cm}+ 9 p_a^3 p_b^3 v_2(p_a) v_2(p_b) \cos\left[2 \Psi_2(p_a) - 2 \Psi_2(p_b)\right] 
+ 3 p_a p_b^5 v_2(p_a) v_2(p_b) \cos\left[2 \Psi_2(p_a) - 2 \Psi_2(p_b)\right] \\
&\hspace{0.45cm}+ \frac{9}{2} p_a^5 p_b v_{2F}^2 v_2(p_a) v_2(p_b) \cos\left[2 \Psi_2(p_a) - 2 \Psi_2(p_b)\right] 
+ \frac{27}{2} p_a^3 p_b^3 v_{2F}^2 v_2(p_a) v_2(p_b) \cos\left[2 \Psi_2(p_a) - 2 \Psi_2(p_b)\right] \\
&\hspace{0.45cm}+ \frac{9}{2} p_a p_b^5 v_{2F}^2 v_2(p_a) v_2(p_b) \cos\left[2 \Psi_2(p_a) - 2 \Psi_2(p_b)\right] 
- 3 p_a^5 p_b v_{2F} v_2(p_b) \cos\left[2 \Psi_2(p_b)\right] \\
&\hspace{0.45cm}- 6 p_a^3 p_b^3 v_{2F} v_2(p_b) \cos\left[2 \Psi_2(p_b)\right] - \frac{3}{4} p_a^5 p_b v_{2F}^3 v_2(p_b) \cos\left[2 \Psi_2(p_b)\right]
 - \frac{3}{2} p_a^3 p_b^3 v_{2F}^3 v_2(p_b) \cos\left[2 \Psi_2(p_b)\right] \\
 &\hspace{0.45cm}+ \frac{3}{4} p_a^5 p_b v_{2F}^2 v_2(p_a) v_2(p_b) \cos\left[2 \Psi_2(p_a) + 2 \Psi_2(p_b)\right]
 + \frac{3}{4} p_a p_b^5 v_{2F}^2 v_2(p_a) v_2(p_b) \cos\left[2 \Psi_2(p_a) + 2 \Psi_2(p_b)\right] \\
 &\hspace{0.45cm}+ p_a^6 v_3(p_a) v_3(p_b) \cos\left[3 \Psi_3(p_a) - 3 \Psi_3(p_b)\right]
 + 9 p_a^4 p_b^2 v_3(p_a) v_3(p_b) \cos\left[3 \Psi_3(p_a) - 3 \Psi_3(p_b)\right] \\
&\hspace{0.45cm}+ 9 p_a^2 p_b^4 v_3(p_a) v_3(p_b) \cos\left[3 \Psi_3(p_a) - 3 \Psi_3(p_b)\right] 
+ p_b^6 v_3(p_a) v_3(p_b) \cos\left[3 \Psi_3(p_a) - 3 \Psi_3(p_b)\right] \\
&\hspace{0.45cm}+ \frac{3}{2} p_a^6 v_{2F}^2 v_3(p_a) v_3(p_b) \cos\left[3 \Psi_3(p_a) - 3 \Psi_3(p_b)\right]
 + \frac{27}{2} p_a^4 p_b^2 v_{2F}^2 v_3(p_a) v_3(p_b) \cos\left[3 \Psi_3(p_a) - 3 \Psi_3(p_b)\right]  \\
&\hspace{0.45cm}+ \frac{27}{2} p_a^2 p_b^4 v_{2F}^2 v_3(p_a) v_3(p_b) \cos\left[3 \Psi_3(p_a) - 3 \Psi_3(p_b)\right]
 + \frac{3}{2} p_b^6 v_{2F}^2 v_3(p_a) v_3(p_b) \cos\left[3 \Psi_3(p_a) - 3 \Psi_3(p_b)\right] \\
&\hspace{0.45cm}- \frac{1}{8} p_a^6 v_{2F}^3 v_3(p_a) v_3(p_b) \cos\left[3 \Psi_3(p_a) + 3 \Psi_3(p_b)\right] 
- \frac{1}{8} p_b^6 v_{2F}^3 v_3(p_a) v_3(p_b) \cos\left[3 \Psi_3(p_a) + 3 \Psi_3(p_b)\right].
\label{B}
\end{aligned}
\end{equation}
The above results are based on the expansion of the exponential term in Eq.~\eqref{f2v3} to the third order, i.e., $\text{exp} \left ( -X \right ) \approx 1-X+\frac{X^{2}}{2}-\frac{X^{3}}{6} $, since the first non-vanishing pure TMC term, i.e., the first term of $B_3$ in Eq.~\eqref{B}, appears at $\frac{X^{3}}{6}$. For the denominator of Eq.~\eqref{r3n}, we only take the first term of its expansion, so the result is $\left ( 2\pi \right ) ^{2} $. For Eq.~\eqref{B}, which also contains a complicated expansion, we apply the same approximation strategy to isolate the dominant terms, yielding:
\begin{equation}
    c_{3}\left \{ 2 \right \}\!\mid _{p_{a}p_{b} }=v_{3}\left ( p_{a}  \right ) v_{3}\left ( p_{b}  \right )  -\frac{p_{a}^{3}p_{a}^{3} }{6(N-2)^{3}  \left \langle p^{2}  \right \rangle _{F}^{3}}
-\frac{v_{2}\left ( p_{a}  \right ) v_{2}\left ( p_{b}  \right )p_{a}p_{b}+v_{3}\left ( p_{a}  \right )v_{3}\left ( p_{b}  \right )p_{a}^{2}+v_{3}\left ( p_{a}  \right ) v_{3}\left ( p_{b}  \right )p_{b}^{2} }
{(N-2)  \left \langle p^{2}  \right \rangle _{F}},
\end{equation}
where the pure TMC term  $-p_{a}^{3}p_{a}^{3} /\ [6(N-2)^{3}  \left \langle p^{2}  \right \rangle _{F}^{3}]$ is the key term that leads to $r_3 > 1$ in the subsequent calculations.
And a numerical comparison between the dominant terms and the full expressions in the experimentally relevant region is presented in Sec. \ref{secB}. \\
In summary, the reason why TMC causes $r_n$ to deviate from 1 is that the pure TMC term plays a decisive role, which scales as $\frac{1}{n!}\frac{p_{a}^{n}p_{b}^{n}  }{\left ( N-2 \right )^n\left \langle p^{2}  \right \rangle_{F}^n   }$, and whose sign alternates as $\left ( -1 \right )^{n}$. This will be elaborated in greater detail in Sec. \ref{secB}. \\ 
For the denominator of $r_3$ in Eq.~\eqref{r2r3}, each factor under the square root corresponds to evaluating Eqs.~\eqref{r3n} and~\eqref{B} with both $p_a$ and $p_b$ uniformly set to $p_a$ or to $p_b$, respectively.

\end{widetext}

\section{Results and discussion}
\subsection{Comparison with CMS data: $r_2$ and $r_3$}

\begin{figure*}[!htb]
    \centering
    \includegraphics[scale=0.7]{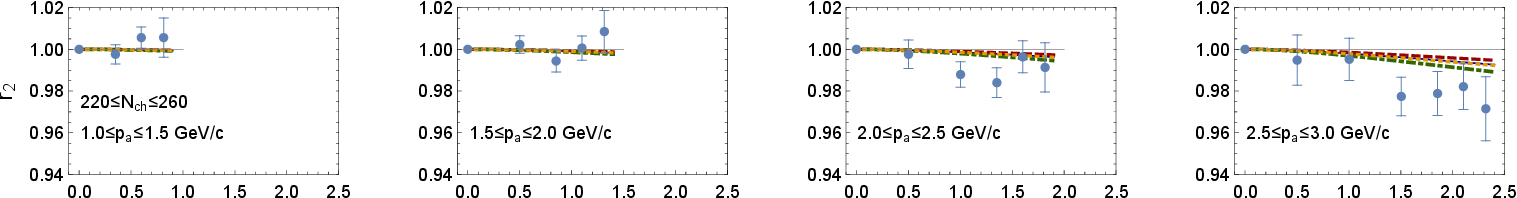}     \includegraphics[scale=0.7]{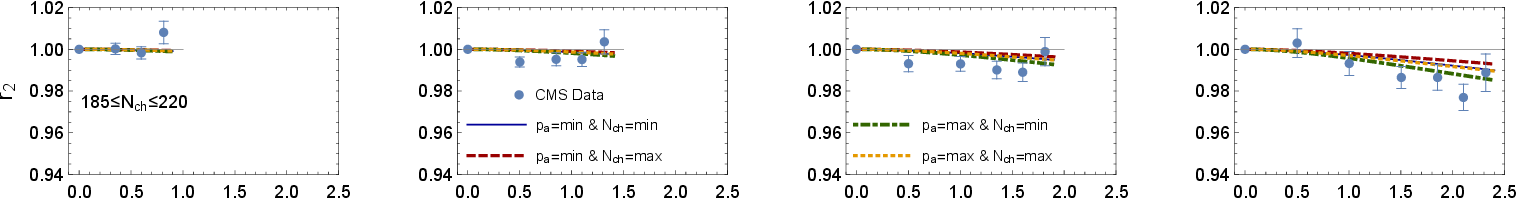}
    \includegraphics[scale=0.7]{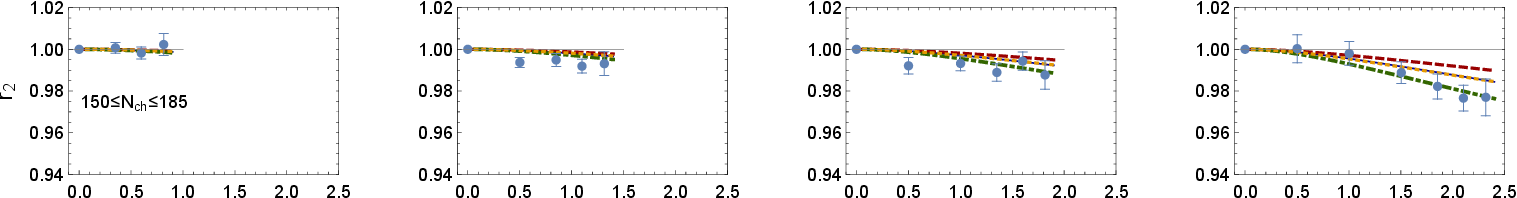}
\includegraphics[scale=0.7]{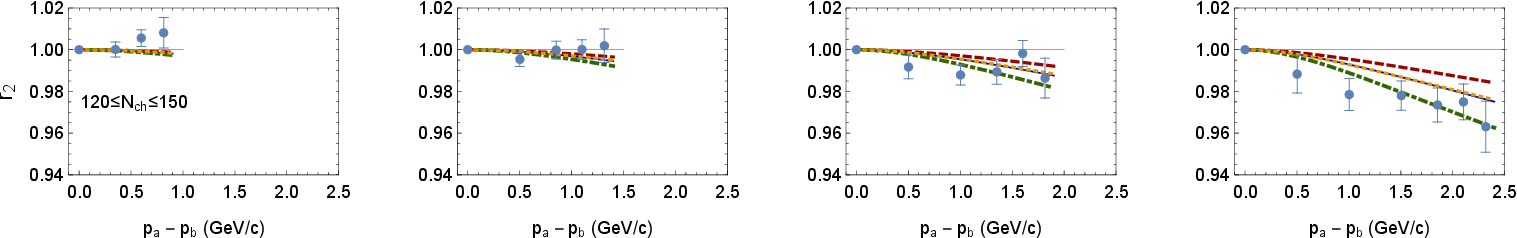}
    \caption{Factorization ratio $r_2$ as a function of $p_{a}-p_{b}$ in four $p_a$ bins (columns) and four $N_{ch}$ ranges (rows), forming a 4$\times$4 array of panels. Curves: TMC calculations. Points: CMS p-Pb data at 5.02 TeV with statistical errors (systematic uncertainties negligible)~\cite{cmsdata}.}
    \label{cos0r2}
\end{figure*}

In Fig.~\ref{cos0r2}, we present the calculated factorization ratio $r_2$ as a function of $p_{a}-p_{b}$ in four $p_a$ bins and four $N_{ch}$ ranges, compared with the CMS p-Pb data at 5.02 TeV. Using the extreme values of the $p_a$ and $N_{\mathrm{ch}}$ ranges from the CMS cuts, we set four parameter combinations for $p_a$ and $N_{ch}$ (with $N = 1.5 N_{ch}$) and calculate the corresponding theoretical curves via Eqs.~\eqref{r2r3} and \eqref{r2n}, yielding four curves in each panel. The $p$-dependence of $v_2$ is extracted from fits to the measured $v_{2}\left \{2\right \} $ and $v_{2}\left \{4\right \} $ under the same experimental conditions \cite{cmsvn} (see Appendix \ref{seca} for details). Our results are practically insensitive to the choice of $v_{2F}$ (see Appendix \ref{secv2F} for details), which we set to $v_{2F}=0.025$.
For $\left \langle p^{2}  \right \rangle_{F} $, we assume
\begin{equation}
    f(p)\propto \text{exp}(-\frac{p}{T} ) ,
\end{equation}
which gives $\left \langle p \right \rangle _{F}=2T $ and $\left \langle p^{2}  \right \rangle _{F}=6T^{2} $. According to Ref.~\cite{meanpt}, the value of $\left \langle p \right \rangle  $ in p-Pb collisions at 5.02 TeV is approximately 0.78 GeV/c for high multiplicities, we then set $\langle p^2 \rangle_F = 0.91$ (GeV/c)$^2$. The correlations between event planes at different $p$ in Eq.~\eqref{A} are approximated by the experimental values shown in the first row of panels of Fig.~\ref{cos0r2}.  
Since high-multiplicity p-Pb events are predominantly flow-driven, our analytical results converge to the form of Eq.~\eqref{flowrn}. Consequently, $r_2$ reduces to $\cos   \left [ 2\Psi _{2} (p_{a} )-2\Psi _{2} (p_{b} ) \right ] $ in this limit (see Appendix \ref{secb} for details). But to highlight the TMC contribution more clearly, we present the results calculated neglecting $p_T$-dependent event plane fluctuations, i.e., setting $\cos  \left [ 2\Psi _{2} (p_{a} )-2\Psi _{2} (p_{b} ) \right ]=1$, as shown in Fig.~\ref{cos0r2}.
In addition, we present the results obtained by considering the $p_T$-dependent event plane fluctuations (see Appendix \ref{secc} for details). With the above parameter set, our calculated $r_2$ is found to be below unity and to decrease with increasing $p_a-p_b$, in good agreement with the CMS data for all studied kinematic cuts. The origin of $r_{2} \le 1$ lies in the TMC effect, which will be demonstrated and explained in Figs.~\ref{c2} and \ref{rnpart} and the accompanying text. 
The factorization breaking effect strengthens progressively with decreasing multiplicity (top to bottom in each column) and increasing momentum (left to right in each row). This is consistent with the characteristic of the TMC effect being more significant at lower multiplicities and higher momenta. Thus, at lower multiplicities and higher momenta, factorization breaking is most pronounced for the green dash-dotted lines, and least for the red dashed ones. The blue solid and orange dotted lines lie between these extremes, essentially overlapping. 
This underscores that the TMC effect must therefore be accounted for in interpreting $r_2$, particularly at lower multiplicities and higher momenta.

\begin{figure*}[!htb]
\centering
    \includegraphics[scale=0.7]{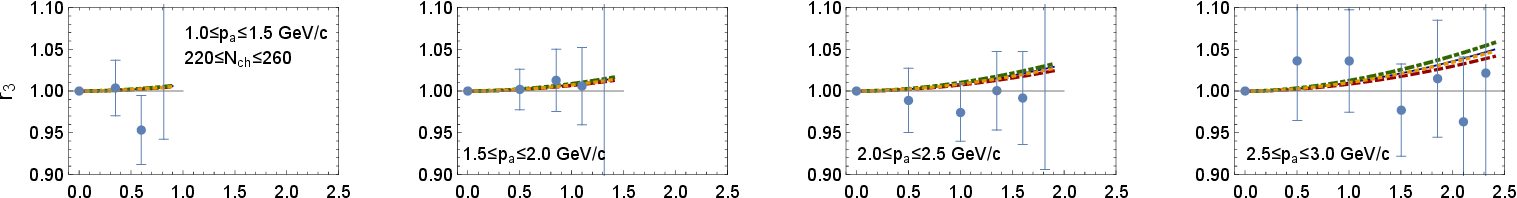}     \includegraphics[scale=0.7]{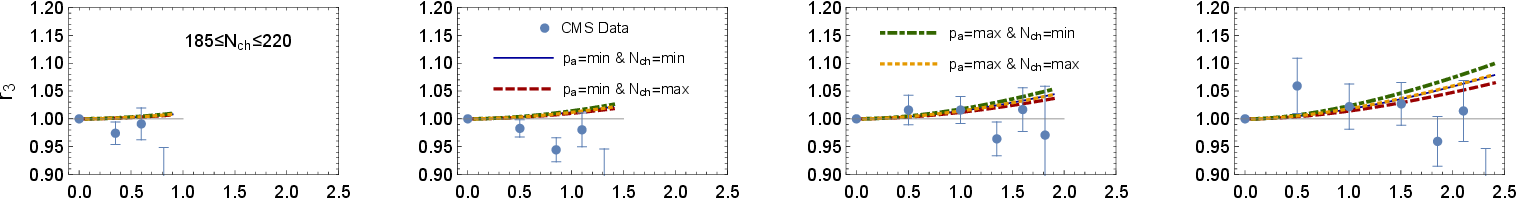}
    \includegraphics[scale=0.7]{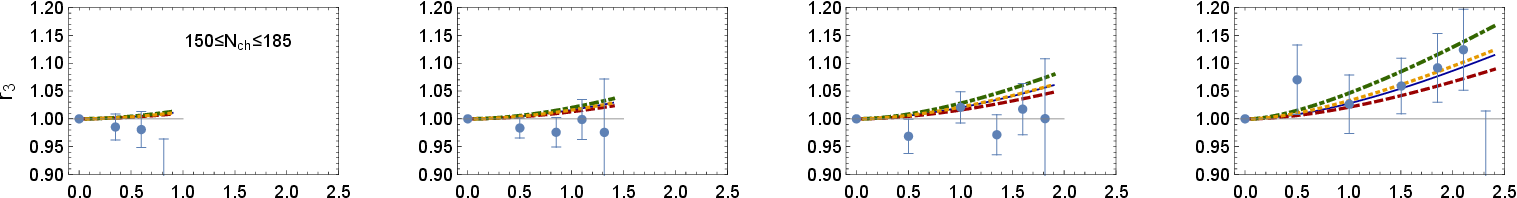}
\includegraphics[scale=0.7]{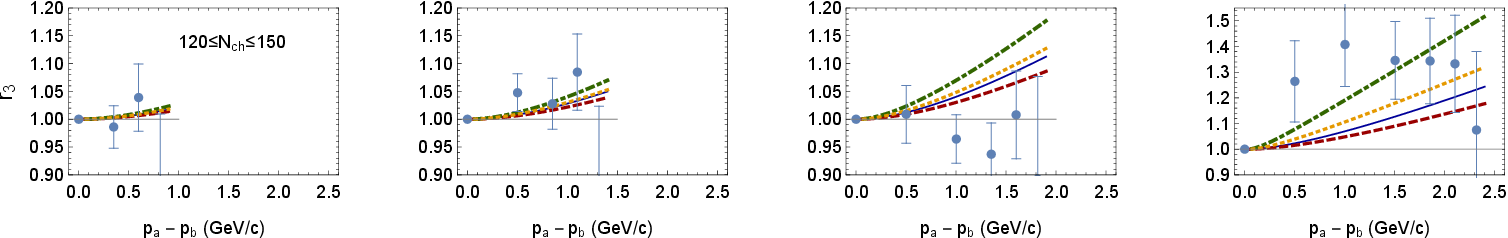}
    \caption{
Factorization ratio $r_3$ as a function of $p_{a}-p_{b}$ in four $p_a$ bins (columns) and four $N_{ch}$ ranges (rows), forming a 4$\times$4 array of panels. Curves: TMC calculations. Points: CMS p-Pb data at 5.02 TeV with statistical errors (systematic uncertainties negligible)~\cite{cmsdata}.     
    }
    \label{cos0r3}
\end{figure*}

In Fig.~\ref{cos0r3}, we present the calculated factorization ratio $r_3$ as a function of $p_{a}-p_{b}$ in four $p_a$ bins and four $N_{ch}$ ranges, compared with the CMS p-Pb data at 5.02 TeV, with all parameter settings consistent with those in Fig.~\ref{cos0r2}. Following the same logic as for $r_2$, $r_3$ should reduce to $\cos[3\Psi_3(p_a) - 3\Psi_3(p_b)]$ in high-multiplicity p-Pb events. However, as the experimental data in Fig.~\ref{cos0r3} (top row) are consistent with unity within uncertainties, we ignore in our analysis the $p_T$-dependent 3rd event-plane correlations in Eq.~\eqref{B}. For the second-order event plane that still appears in $r_3$, its setting method is consistent with that of $r_2$ in Fig.~\ref{cos0r2}. Similarly, we present the results calculated neglecting $p_T$-dependent event plane fluctuations, i.e., with $\cos  \left [ 2\Psi _{2} (p_{a} )-2\Psi _{2} (p_{b} ) \right ]=1$, as shown in Fig.~\ref{cos0r3}.
Furthermore, we also present the results of $r_3$ obtained  by considering the $p_T$-dependent second-order event plane fluctuations and find that its effect is negligible (see Appendix \ref{secc} for details). We observe that $r_3 > 1$ emerges in our analysis, in sharp contrast to $r_2 < 1 $ in Fig.~\ref{cos0r2}. This is attributed to the TMC effect which is most evident at lower multiplicities and higher momenta.
Figures~\ref{rnpart} and ~\ref{c3} (and accompanying text) will reveal the mechanism by which the TMC effect drives $r_3$ above unity. Although $r_3$ and $r_2$ deviate from unity in opposite directions, the magnitude of deviation for both increases monotonically with $p_a-p_b$.
Notably, $r_3$ exhibits a much stronger dependence on multiplicity than $r_2$. If taking the last-column curves (e.g. the orange dotted lines) in Figs.~\ref{cos0r2} and \ref{cos0r3} as an example, we observe that $r_2$ varies by only about 2\% from top to bottom, whereas $r_3$ changes by about 20\%. Our TMC-based calculations reproduce the experimental data within uncertainties, capturing both $r_3 > 1$ and its trends with $p_a$ and multiplicity. These findings support TMC as the mechanism responsible for $r_3 > 1$ in lower-multiplicity p+Pb events.

\subsection{The sign rule of $r_n$}
\label{secB}

\begin{figure*}[!htb]
\centering
    \includegraphics[scale=0.8]{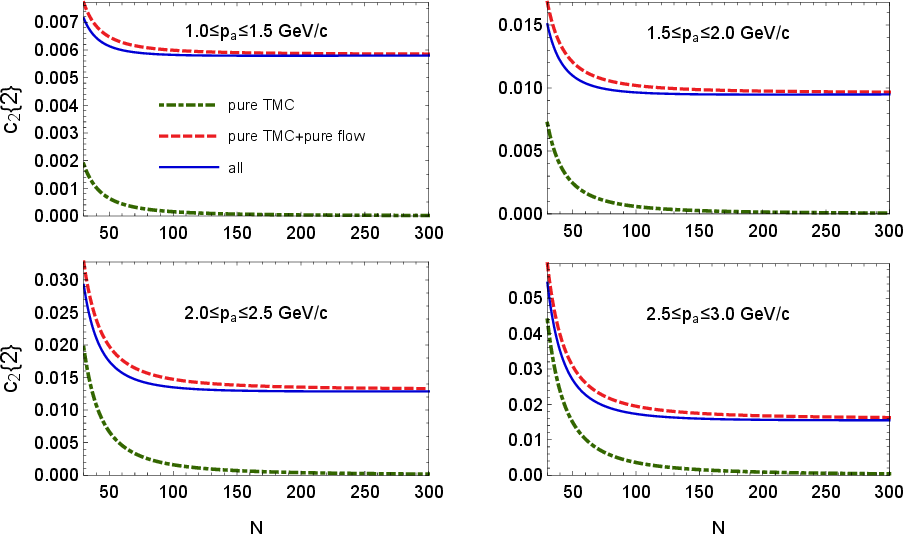}
    \caption{$c_{2}\left \{ 2 \right \} $ from ``pure TMC'', ``pure TMC+pure flow'', and ``pure TMC+pure flow+interplay'' (labeled as ``all") as a function of the number of particles $N$ for four $p_a$ bins.}
    \label{c2}
\end{figure*}

\begin{figure*}[!htb]
\centering
    \includegraphics[scale=0.7]{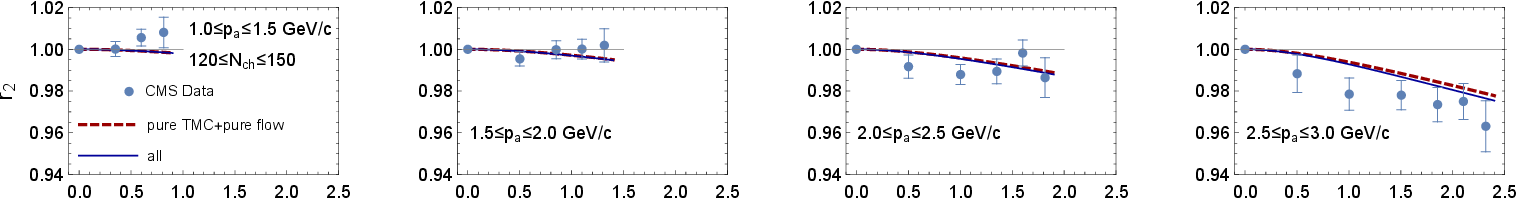}
    \includegraphics[scale=0.7]{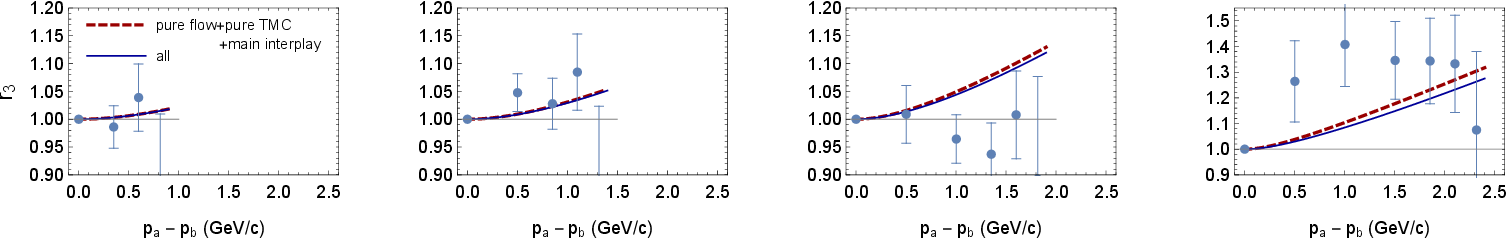}
    \caption{
Factorization ratio $r_2$ (upper row) and $r_3$ (lower row) as a function of $p_{a}-p_{b}$ in four $p_{a}$ bins (columns) for $120\le N_{ch}\le 150 $. Curves: calculations from TMC and flow (proxy and all). Points: CMS p-Pb data at 5.02 TeV with statistical errors (systematic uncertainties negligible)~\cite{cmsdata}. 
}
    \label{rnpart}
\end{figure*}

In this part, we will further explain how TMC affects $r_2$ and $r_3$. Essentially, $r_n$ reflects the ratio of the azimuthal correlation between particles with different momenta to that between particles with the same momenta, i.e., the relationship between $c_{n}\left \{ 2 \right \}\!\mid _{p_{a},p_{b} }$, $c_{n}\left \{ 2 \right \}\!\mid _{p_{a} }$, and $c_{n}\left \{ 2 \right \}\!\mid _{p_{b} }$, $r_{n} =\frac{c_{n}\left \{ 2 \right \}\!\mid _{p_{a},p_{b} }}
{\sqrt{\left ( c_{n}\left \{ 2 \right \}\!\mid _{p_{a} } \right )\times \left ( c_{n}\left \{ 2 \right \}\!\mid _{p_{b} } \right )  } } $. As can be seen from Eqs.~\eqref{A} and~\eqref{B}, the analytical result for $c_{n}\left \{ 2 \right \}\!\mid _{p_{a},p_{b} }$ is quite complex. Therefore, we need to extract the most critical terms from them as the basis for subsequent analyses. To isolate the influence of TMC, we neglect event-plane fluctuations by assuming $\Psi_n(p_a) = \Psi_n(p_b)$. We have verified that relaxing this assumption does not alter our conclusions, see Appendix \ref{secc} for details. The key terms in $c_2\{2\}$ are given by: \\
Type I (pure TMC):
\begin{equation}
    c_{2}\left \{ 2 \right \}\!\mid _{p_{a} }=
  \frac{p_{a}^{4}  }{2(N-2)^{2}  \left \langle p^{2}  \right \rangle _{F}^{2}} ,
  \label{c2t}
\end{equation}
Type II (pure flow + pure TMC):
\begin{equation}
    c_{2}\left \{ 2 \right \}\!\mid _{p_{a} }=v_{2}\left ( p_{a}  \right )^{2}+
  \frac{p_{a}^{4}  }{2(N-2)^{2}  \left \langle p^{2}  \right \rangle _{F}^{2}}.
  \label{c2ft}
\end{equation}
In Fig.~\ref{c2} we compare the results based on Eqs.~\eqref{c2t} and~\eqref{c2ft} with the complete results (labeled as ``all''). Here, parameter $p_a$ is taken as the median values within their respective range, while the selection of other parameters is consistent with those in Fig.~\ref{cos0r2}.
Figure~\ref{c2} demonstrates that the “pure TMC + pure flow” contribution accounts for the most of the complete result for all four momentum bins. Therefore, we use the concise ``pure TMC + pure flow” result as an efficient proxy for the complete result in the following analysis. Then, $r_2$ can be rewritten as follows,
    \begin{equation}
    r_{2} =\frac{v_{2}\left ( p_{a}  \right )v_{2}\left ( p_{b}  \right )+
  \frac{p_{a}^{2}p_{b}^{2}  }{2(N-2)^{2}  \left \langle p^{2}  \right \rangle _{F}^{2}}}
{\sqrt{(v_{2}\left ( p_{a}  \right )^{2}+
  \frac{p_{a}^{4}  }{2(N-2)^{2}  \left \langle p^{2}  \right \rangle _{F}^{2}})
\times(v_{2}\left ( p_{b}  \right )^{2}+
  \frac{p_{b}^{4}  }{2(N-2)^{2}  \left \langle p^{2}  \right \rangle _{F}^{2}})} } .
  \label{r2main}
\end{equation}
A comparison between Eq.~\eqref{r2main} and the complete result is shown in the first row of Fig.~\ref{rnpart}. We take the results in the low-multiplicity region ($120 \le N_{\text{ch}} \le 150$) as an example, since the TMC effect is relatively prominent here, as illustrated in Fig.~\ref{cos0r2}.
The excellent agreement between the two results validates the use of this approximation. By the Cauchy-Schwarz inequality, the expression in Eq.~\eqref{r2main} is always less than or equal to 1. This indicates that on top of the pure flow contribution, TMC introduces additional correlations between the azimuthal angles of particles.

\begin{figure*}[!htb]
\centering
\includegraphics[scale=0.8]
{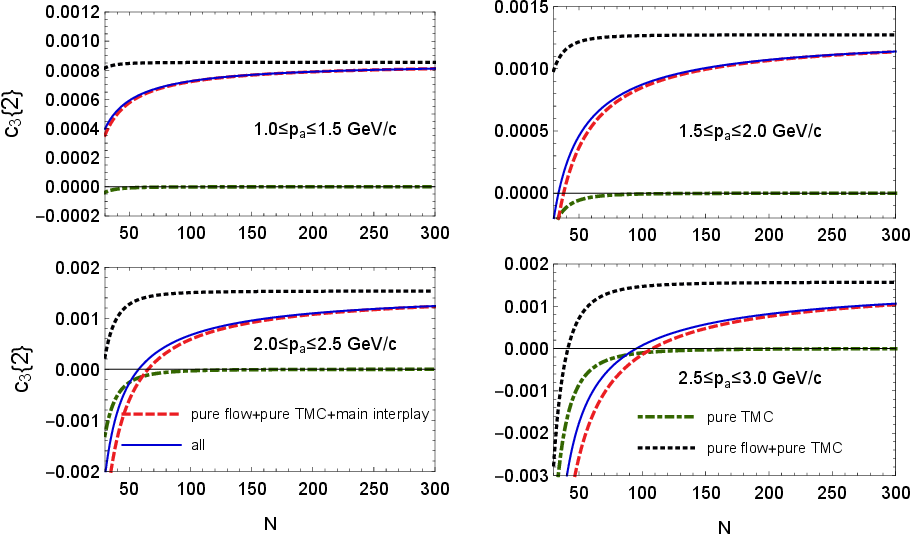}
    \caption{$c_{3}\left \{ 2 \right \}$ from ``pure TMC'', ``pure flow+pure TMC'', ``pure flow+pure TMC+main interplay'', and ``pure flow+pure TMC+interplay''(labeled ``all") as a function of the number of particles $N$ for four $p_a$ bins.}
    \label{c3}
    \label{c3}
\end{figure*}

In a similar way, we obtain $c_{3}\left \{ 2 \right \}\!\mid _{p_{a} }$ and the simplified expression for $r_{3}$:\\
Type I (pure TMC):
\begin{equation}
    c_{3}\left \{ 2 \right \}\!\mid _{p_{a} }=
-\frac{p_{a}^{6} }{6(N-2)^{3}  \left \langle p^{2}  \right \rangle _{F}^{3}},
\label{c3t}
\end{equation}
Type II (pure flow + pure TMC):
\begin{equation}
    c_{3}\left \{ 2 \right \}\!\mid _{p_{a} }=v_{3}\left ( p_{a}  \right ) ^{2}  
-\frac{p_{a}^{6} }{6(N-2)^{3}  \left \langle p^{2}  \right \rangle _{F}^{3}},
\label{c3ft}
\end{equation}
\begin{widetext}
    Type III (pure flow + pure TMC + main interplay):
\begin{equation}
    c_{3}\left \{ 2 \right \}\!\mid _{p_{a} }=v_{3}\left ( p_{a}  \right ) ^{2}  -\frac{p_{a}^{6} }{6(N-2)^{3}  \left \langle p^{2}  \right \rangle _{F}^{3}}
-\frac{v_{2}\left ( p_{a}  \right ) ^{2}p_{a}^{2}+2v_{3}\left ( p_{a}  \right ) ^{2}p_{a}^{2} }
{(N-2)  \left \langle p^{2}  \right \rangle _{F}}.
\label{c3fm}
\end{equation}
\end{widetext}
Fig.~\ref{c3} compares the results based on Eqs.~\eqref{c3t}, \eqref{c3ft}, and \eqref{c3fm} with the complete result, showing that the ``pure flow + pure TMC  + main interplay" closely reproduces the complete result across broad momentum and multiplicity ranges. Therefore, Eq.~\eqref{c3fm} will be used as an efficient proxy for the complete result. Then $r_3$ can be rewritten as follows:
\begin{widetext}
    \begin{equation}
    r_{3} =\frac{v_{3}\left ( p_{a}  \right )v_{3}\left ( p_{b}  \right )-\frac{p_{a}^{3}p_{b}^{3} }{6(N-2)^{3}  \left \langle p^{2}  \right \rangle _{F}^{3}}  
-\frac{v_{2}\left ( p_{a}  \right )v_{2}\left ( p_{b}  \right ) p_{a}p_{b}+v_{3}\left ( p_{a} \right )v_{3}\left ( p_{b} \right ) \left (p_{a}^{2}+ p_{b}^{2} \right )  }
{(N-2)  \left \langle p^{2}  \right \rangle _{F}}}
{\sqrt{(v_{3}\left ( p_{a}  \right ) ^{2}-\frac{p_{a}^{6} }{6(N-2)^{3}  \left \langle p^{2}  \right \rangle _{F}^{3}}  
-\frac{v_{2}\left ( p_{a}  \right ) ^{2}p_{a}^{2}+2v_{3}\left ( p_{a}  \right ) ^{2}p_{a}^{2} }
{(N-2)  \left \langle p^{2}  \right \rangle _{F}})
\times(v_{3}\left ( p_{b}  \right ) ^{2}-\frac{p_{b}^{6} }{6(N-2)^{3}  \left \langle p^{2}  \right \rangle _{F}^{3}}  
-\frac{v_{2}\left ( p_{b}  \right ) ^{2}p_{b}^{2}+2v_{3}\left ( p_{b}  \right ) ^{2}p_{b}^{2} }
{(N-2)  \left \langle p^{2}  \right \rangle _{F}})} } .
\label{r3main}
\end{equation}
\end{widetext}
The second row of Fig.~\ref{rnpart} compares the results from Eq.~\eqref{r3main} with the complete results, demonstrating their close agreement and thereby validating this substitution. That $r_{3}\ge 1$ can be easily proven by a simple mathematical rearrangement of Eq.~\eqref{r3main}.
Physically, pure TMC and the interplay between TMC and flow weakens the 3rd order of azimuthal correlation relative to the pure flow baseline. This weakening is formally shown to be less pronounced for particles with differing momenta than for those with equal momenta. Consequently, the numerator of Eq.~\eqref{r3main} becomes larger than the denominator, resulting in $r_3\ge1$.

Next, based on the above analysis, we will summarize the general regularity regarding whether TMC causes $r_n$ to be greater than 1 or less than 1. For the two-particle correlation cumulants of any order, namely $c_{n}\left \{ 2 \right \} $, we can divide their analytical results into three categories: pure TMC, pure flow, and their interplay term. The three categories contribute hierarchically: pure TMC reflects the variation trend of $c_{n}\left \{ 2 \right \} $ with multiplicity; pure flow adjusts the sign and order of magnitude of $c_{n}\left \{ 2 \right \} $ on the basis of pure TMC; and the interplay term further fine-tunes the numerical value based on the previous two. 
Therefore, the pure TMC and pure flow contributions capture the essential trends and order of magnitude of $c_{n}\left \{ 2 \right \} $. This justifies our subsequent analysis of $r_n$, which will be framed in terms of these two dominant components. For $c_{n}\left \{ 2 \right \}\!\mid _{p_{a},p_{b} }$, the specific form of pure flow is always $v_{n}\left ( p_{a}  \right )v_{n}\left ( p_{b}  \right ) $. Next, we derive the general formula for pure TMC. That is, without considering $v_n$, the calculation of $c_{n}\left \{ 2 \right \}\!\mid _{p_{a},p_{b} }$ is as follows:
\begin{equation}
    c_{n}\left \{ 2 \right \}\!\mid _{p_{a},p_{b} }=
\left \langle
 e^{in\left ( \phi _{1}^{a}-\phi _{2}^{b}  \right ) }  \right \rangle=
\frac{\int_{0}^{2\pi}e^{in\left ( \phi _{1}^{a}-\phi _{2}^{b}  \right ) }\text{exp}\left ( X \right)
 d \phi _{1}^{a} d \phi _{2}^{b} }
{\int_{0}^{2\pi}\text{exp}\left ( X \right)
 d \phi _{1}^{a} d \phi _{2}^{b}} ,
\end{equation}
where, according to Eq.~\eqref{f2s}, the form of $X$ is as follows:
\begin{equation}
    X=-\frac{2p_ap_b\cos \left ( \phi _{1} ^{a}-\phi_{2} ^{b}  \right ) }
{\left ( N-2 \right )\left \langle p^{2}  \right \rangle_{F}  }  .
\label{X}
\end{equation}
We still perform a Taylor expansion on the exponential term in Eq.~\eqref{X}, i.e., $\text{exp}\left ( X \right ) = 1+X+\frac{X^{2} }{2}+\frac{X^{3} }{6} +\cdots$, expanding until the first non-zero result appears. However, for the denominator, we only take the first term of its expansion, so the result is $\left ( 2\pi \right ) ^{2} $. Therefore, the general formula for pure TMC is $\left ( -1 \right )^{n}\frac{1}{n!}\frac{p_{a}^{n}p_{b}^{n}  }{\left ( N-2 \right )^n\left \langle p^{2}  \right \rangle_{F}^n   } $. 
\begin{widetext}
Then, $r_n$ can be written in the following form (this serves as a useful approximation that, while not exact, reproduces the key features.):

    \begin{equation}
    r_{n} =\frac{v_{n}\left ( p_{a}  \right )v_{n}\left ( p_{b}  \right )+
 \left ( -1 \right )^{n}\frac{1}{n!}\frac{p_{a}^{n}p_{b}^{n}  }{\left ( N-2 \right )^n\left \langle p^{2}  \right \rangle_{F} ^n  }}
{\sqrt{\left (  v_{n}\left ( p_{a}  \right )^{2}+
  \left ( -1 \right )^{n}\frac{1}{n!}\frac{p_{a}^{2n} }{\left ( N-2 \right )^n \left \langle p^{2}  \right \rangle_{F} ^n   }\right )\times\left (  v_{n}\left ( p_{b}  \right )^{2}+
  \left ( -1 \right )^{n}\frac{1}{n!}\frac{p_{b}^{2n} }{\left ( N-2 \right )^n \left \langle p^{2}  \right \rangle_{F} ^n   } \right )  } } .
  \label{rnmain}
\end{equation}
From Eq.~\eqref{rnmain}, it can be seen that without the TMC effect, $r_{n} =\frac{v_{n}\left ( p_{a}  \right )v_{n}\left ( p_{b}  \right )}
{\sqrt{v_{n}\left ( p_{a}  \right )^{2} v_{n}\left ( p_{b}  \right )^{2} } } =1$. However, the introduction of the TMC effect will change the azimuthal correlation of two particles. Next, squaring both the numerator and denominator of Eq.~\eqref{rnmain} allows for a direct observation of the TMC contribution.
    The result of squaring the numerator of Eq.~\eqref{rnmain} is as follows:
\begin{align}
   U&= v_{n}\left ( p_{a}  \right )^{2} v_{n}\left ( p_{b}  \right )^{2}
+\frac{1}{\left ( n! \right ) ^{2} }\frac{p_{a}^{2n}p_{b}^{2n}}{\left ( N-2 \right )^{2n}\left\langle p^{2}\right\rangle_{F}^{2n} }
+2v_{n}\left ( p_{a}  \right ) v_{n}\left ( p_{b}  \right )
\left ( -1 \right )^{n}\frac{1}{n!}\frac{p_{a}^{n}p_{b}^{n}  }{\left ( N-2 \right )^n\left \langle p^{2}  \right \rangle_{F}^n   } \notag \\
&=U_1+U_2+U_3,
\label{U}
\end{align}
where the sum of the second and third terms ($U_2+U_3$) in Eq.~\eqref{U} are the changes in the azimuthal correlation of two particles with different momenta caused by the introduction of TMC. 
The result of squaring the denominator of Eq.~\eqref{rnmain} is as follows:
\begin{align}
D&=    v_{n}\left ( p_{a}  \right )^{2} v_{n}\left ( p_{b}  \right )^{2}
+\frac{1}{\left ( n! \right ) ^{2} }\frac{p_{a}^{2n}p_{b}^{2n}}{\left ( N-2 \right )^{2n}\left\langle p^{2}\right\rangle_{F}^{2n} }
+v_{n}\left ( p_{a}  \right )^{2}\left ( -1 \right )^{n}\frac{1}{n!}\frac{p_{b}^{2n}  }{\left ( N-2 \right )^n\left \langle p^{2}\right \rangle_{F}^n}
+v_{n}\left ( p_{b}  \right )^{2}\left ( -1 \right )^{n}\frac{1}{n!}\frac{p_{a}^{2n}  }{\left ( N-2 \right )^n\left \langle p^{2}\right \rangle_{F}^n} \notag \\
&=D_1+D_2+D_3+D_4,
\label{D}
\end{align}
\end{widetext}
where the sum of the second, third, and fourth terms ($D_2+D_3+D_4$) in Eq.~\eqref{D} are the changes in the azimuthal correlation of two particles with the same momentum caused by the introduction of TMC. Since $D_2 = U_2$, the comparison of the TMC effect between particles with different momenta and those with the same momentum reduces to comparing $U_3$ to $D_3 + D_4$. 
According to the mean inequality, $A+B\ge2\sqrt{AB} $ (where $A\ge 0$ and $B\ge 0$), we have:
\begin{equation}
    \begin{cases}
D_3+D_4\ge 2\sqrt{D_3D_4}=U_3,   & \text{ if } n=2k,k\in \mathbb{N}^+, \\
D_3+D_4\le U_3,  & \text{ if } n=2k+1,k\in \mathbb{N^+}.
\end{cases}
\end{equation}
Thus, for the factorization ratio $r_n$ we have such a conjecture:
\begin{equation}
     \begin{cases}
r_n\le1,   & \text{ if } n=2k,k\in \mathbb{N}^+, \\
r_n\ge1,  & \text{ if } n=2k+1,k\in \mathbb{N}^+.
\end{cases}   
\end{equation}
This is because TMC has a weaker influence on azimuthal correlations for particle pairs with different momenta than for those with identical momenta, i.e., $|U_3| \le |D_3 + D_4|$.  
Consequently, the TMC causes $r_n$ to deviate from unity; the sign of this deviation (whether $r_n > 1$ or $r_n < 1$) is determined by whether the pure TMC contributes positively or negatively to the two-particle azimuthal correlations of different orders.  
We find that the sign of the pure TMC contribution to $c_{n}\left \{ 2 \right \}  $ alternates with the harmonic order $n$ as $(-1)^n$. Note that this method serves only to determine the sign of the deviation of $r_n$ from unity, and cannot be used as an accurate numerical replacement. For example, for $r_3$, an accurate replacement of the complete result must also take into account the impact of the interplay. Additionally, we have also verified $r_4$, and the complete result is indeed less than 1.

While the experimental analysis employs an $\eta$ gap to suppress short-range non-flow correlations, our calculation contains no $\eta$ dependence. Therefore, an $\eta$-dependent cut would not affect our results. Jet-related effects are absent by construction in our calculation, whereas they are present in experimental measurements and motivate the use of an $\eta$ gap. Consequently, the more effectively the jet-induced short-range non-flow contributions are suppressed experimentally, the closer the remaining non-flow contributions are expected to approach the pure TMC effect. However, the experimental cut $|\Delta\eta|>2$ cannot completely eliminate jet contributions, which may introduce some bias when comparing our results with the experimental measurements obtained using $|\Delta\eta|>2$. Furthermore, since di-jet correlations contribute positively to $c_2\{2\}$ and negatively to $c_3\{2\}$, the sign of the deviation of $r_n$ from unity alone does not uniquely identify the dominant non-flow source. Nevertheless, a comparison of the numerical values and their trends suggests that TMC plays an essential role, as discussed in Appendix \ref{secjet}.

\section{Conclusions} 

In this paper, we calculate factorization ratios $r_2$ and $r_3$ of two-particle azimuthal correlations as a function of $p_{a}-p_{b}$ under the constraint of transverse momentum conservation. We observe that the TMC effect causes both $r_2$ and $r_3$ to deviate from 1, with the deviation increasing as the momentum difference between particles grows.  
Physically, the TMC alters azimuthal two-particle correlations relative to the pure flow baseline, applying a smaller correction to pairs with differing momenta than to those with the same momentum. By introducing a positive correlation to $c_{2}\left \{ 2 \right \}  $ but a negative one to $c_{3}\left \{ 2 \right \}  $, the TMC leads to the conclusion that the deviation $r_n - 1$ follows $\left ( - 1 \right )^{n+1}  $, being negative for even and positive for odd harmonic orders $n$.
In addition, it can also be seen that our analytical results for $r_2$ and $r_3$ both show the most significant deviation from 1 at low multiplicities and high momenta, which reflects the characteristics of TMC. 
Our analytical results are in good agreement with the CMS p-Pb data for both $r_2$ and $r_3$ within errors. Therefore, this kinematic effect must be subtracted or accounted for before $r_n$ can be used as a reliable probe of genuine flow fluctuations.
Finally, we find that within the TMC framework, the factorization ratio obeys $r_n < 1$ for even $n$ and $r_n > 1$ for odd $n$. This stems from the sign-alternating pure TMC contribution to $c_{n}\left \{ 2 \right \}  $, whose sign is given by $(-1)^n$. 
A promising direction for future work is to generalize the framework and include the longitudinal momentum conservation. This would enable a comprehensive study of its role in driving longitudinal decorrelation, ultimately clarifying the emergent correlation dynamics in the full phase space of small systems.

\setcounter{section}{0} 
\section*{APPENDIX}
\makeatletter
\@addtoreset{subsection}{section}  
\makeatother
\renewcommand{\thesubsection}{\Alph{subsection}} 
\setcounter{subsection}{0}

\subsection{The effect of $v_3$ on $r_2$}
\label{secwithv3}
\renewcommand{\theequation}{A.\arabic{equation}}
\renewcommand{\thefigure}{A.\arabic{figure}} 
\renewcommand{\thetable}{A.\arabic{table}} 

\setcounter{equation}{0}  
\setcounter{figure}{0}     
\setcounter{table}{0}
In this section, we show that $v_3$ contributes negligibly to $c_2\{2\}$ and thus has no impact on $r_2$. When performing the TMC calculation, if only the effect of $v_2$ is considered, the expression for $c_2\{2\}$ is as follows:
\begin{equation}
    c_2\{2\}\big|_p = v_2^2 - \frac{ p ^2 \left(2 v_2^2 - v_2v_{2F}\right)}{(N-2)\langle p^2\rangle_F} + \frac{p^4}{2(N-2)^2\langle p^2\rangle_F^2}.
    \label{c2_v2}
\end{equation}
Based on Eq.~\eqref{c2_v2}, with $v_3$ further included, the expression becomes:
\begin{equation}
    c_2\{2\}\big|_p = v_2^2 - \frac{ p ^2 \left(2 v_2^2 - v_2v_{2F}+v_3^2\right)}{(N-2)\langle p^2\rangle_F} + \frac{p^4}{2(N-2)^2\langle p^2\rangle_F^2},
    \label{c2_v2v3}
\end{equation}
where the $v_3$ term appears only in the interplay term, rather than in the dominant first pure flow term and the last pure TMC term. Then we plot $c_2\{2\}$ from Eqs.~\eqref{c2_v2} and~\eqref{c2_v2v3} as a function of $N$ in the four experimentally given $p$ bins for comparison in Fig.~\ref{c2_compare}, where for simplicity $p$ in the formulas is taken as the midpoint of each $p$ bin. Fig.~\ref{c2_compare} clearly shows that including or excluding $v_3$ does not affect $c_2\{2\}$, which is why it has no impact on $r_2$.

\begin{figure*}[!htb]
    \centering
    \includegraphics[scale=0.7]{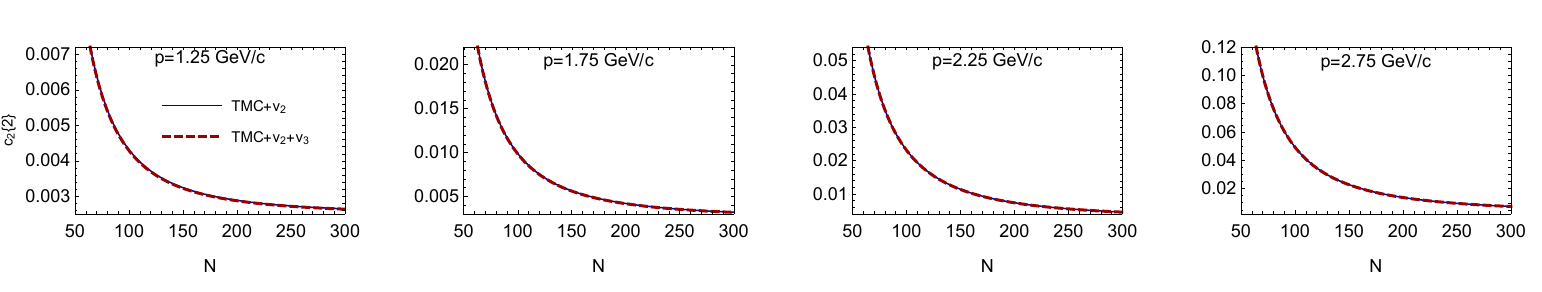}     
    \caption{$c_2\{2\}$ from Eqs.~\eqref{c2_v2} and~\eqref{c2_v2v3} as a function of $N$ for various $p$.}
    \label{c2_compare}
\end{figure*}

\subsection{$v_n(p)$}
\label{seca}
\renewcommand{\theequation}{B.\arabic{equation}}
\renewcommand{\thefigure}{B.\arabic{figure}} 
\renewcommand{\thetable}{B.\arabic{table}} 

\setcounter{equation}{0}  
\setcounter{figure}{0}     
\setcounter{table}{0}
To compare our results with the data, we need to determine the dependence of elliptic flow parameter $v_2$ on transverse momentum $p$. Since CMS has measured the data of $v_{2} \left \{ 2 \right \} $ and $v_{2} \left \{ 4 \right \} $ as functions of $p$ under different multiplicities $N_{ch} $ in p-Pb collisions at 5.02 TeV, we will determine $v_{2} \left ( p \right )$  based on them \cite{cmsvn}. 
Fluctuations in the elliptic flow $v_2$ are driven by event-by-event differences in the initial geometry, including impact parameter and participant nucleon positions.
The fluctuation of $v_2$ can be defined as follows \cite{fluctuation},
\begin{equation}
    \sigma ^{2} =\left \langle v_{2}^{2}  \right \rangle
-\left \langle v_{2}  \right \rangle ^{2}  .
\label{eqA1}
\end{equation} 
We consider how fluctuations affect the flow observable $f(v_2)$. When the function $f\left ( v_{2}  \right )$ is expanded in a series around the average value of the collective flow $\left \langle v_{2}  \right \rangle $  up to the second-order term, the average value of the collective flow observable $f\left ( v_{2}  \right )$, denoted as $\left \langle f\left ( v_{2}  \right )  \right \rangle$, is:
\begin{equation}
    \left \langle f\left ( v_{2}  \right )  \right \rangle =
f\left ( \left \langle v_{2} \right \rangle \right )  +\frac{\sigma ^{2} }{2} 
f''{\left ( \left \langle v_{2} \right \rangle \right )}  .
\label{eqA2}
\end{equation}
Therefore, based on Eqs.~\eqref{eqA1} and ~\eqref{eqA2}, we can respectively derive the collective flows $v_{2} \left \{ 2 \right \} $ and $v_{2} \left \{ 4 \right \} $ measured via two-particle correlations and four-particle correlations:
\begin{align}
    v_{2} \left \{ 2 \right \}^{2} &=\left \langle v_{2}^{2}  \right \rangle=
\left \langle v_{2}  \right \rangle ^{2} +\sigma ^{2} ,\label{eqA3} \\
v_{2} \left \{ 4 \right \}^{2}& =\left (  2\left \langle v_{2}^{2} \right \rangle^{2}
- \left \langle v_{2}^{4}  \right \rangle  \right )^{\frac{1}{2} } \approx 
\left \langle v_{2}  \right \rangle ^{2} -\sigma ^{2} .\label{eqA4}
\end{align}
By combining Eqs.~\eqref{eqA3} and ~\eqref{eqA4}, the fluctuations can be eliminated to obtain $\left \langle v_{2}  \right \rangle $:
\begin{equation}
    \left \langle v_{2}  \right \rangle=\sqrt{\frac{v_{2 }\left \{ 2 \right \}^{2} +v_{2 }\left \{ 4 \right \}^{2}  }
{2} }  .
\end{equation}
We perform a polynomial fit to the experimental data. The specific fitting formulas are given in Table~\ref{tab} and a comparison between the fitting results and the data is shown in Fig.~\ref{vn}. Here, under the same experimental conditions for triangular flow $v_3$, only two-particle correlation data are available, and the value in parentheses is the coefficient of determination, $R^2$.

\begin{table*}[!htb]
  \centering
  \setlength{\tabcolsep}{5pt}
  \caption{Polynomial fits to the $p_T$-dependence of $v_2\{2\}$, $v_2\{4\}$, and $v_3\{2\}$ in four $N_{\text{ch}}$ ranges for p-Pb collisions at 5.02 TeV.}
  \begin{tabular}{c c c c}
    \hline
    \hline
    $N_{ch}$ & $v_2\{2\}$ & $v_2\{4\}$ & $v_3\{2\}$ \\
    \hline 
    $220\le N_{\text{ch}}\le 260$ & $0.0832p-0.0105p^2 \hspace{0.1cm}(0.991)$ & $0.062p-0.008p^2-0.00064p^3+0.0001p^4\hspace{0.1cm}(0.998)$ & $0.036p-0.0056p^2\hspace{0.1cm}(0.943)$  \\
    $185\le N_{\text{ch}}\le 220$ & $0.0827p-0.0101p^2\hspace{0.1cm}(0.993)$ &$0.054p+0.003p^2-0.005p^3+0.0005p^4\hspace{0.1cm}(0.998)$ & $0.034p-0.0055p^2\hspace{0.1cm}(0.943)$ \\
    $150\le N_{\text{ch}}\le 185$ & $0.0823p-0.01p^2\hspace{0.22cm}(0.99)$ & $0.061p+0.001p^2-0.005p^3+0.0006p^4\hspace{0.1cm}(0.996)$& $0.035p-0.0064p^2\hspace{0.1cm}(0.977)$  \\
    $120\le N_{\text{ch}}\le 150$ & $0.0818p-0.0097p^2\hspace{0.1cm}(0.991)$ & $0.057p+0.002p^2-0.006p^3+0.0006p^4\hspace{0.1cm}(0.997)$ & $0.03p-0.0059p^2\hspace{0.1cm}(0.944)$  \\
    \hline 
    \hline 
  \end{tabular}
  \label{tab}
\end{table*}

\begin{figure*}[!htb]
\centering
    \includegraphics[scale=0.7]{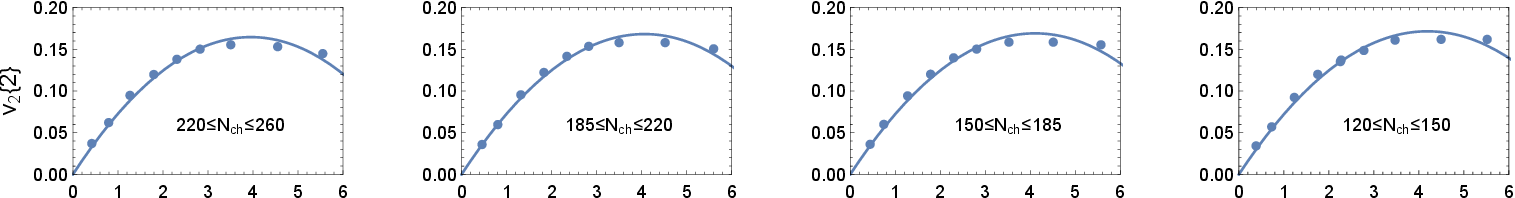}
    \includegraphics[scale=0.7]{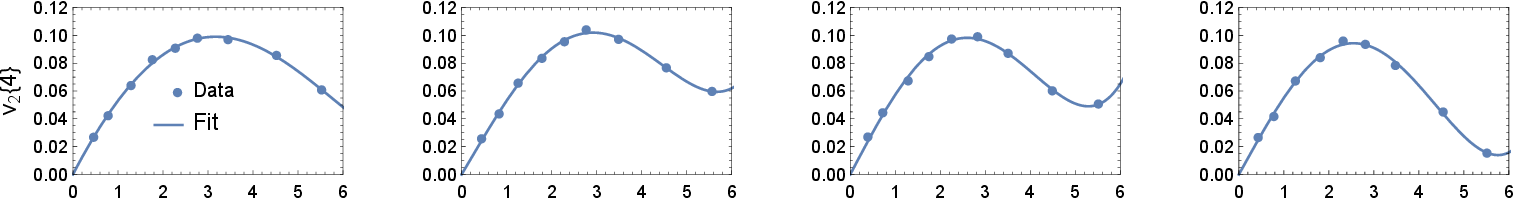}
    \includegraphics[scale=0.7]{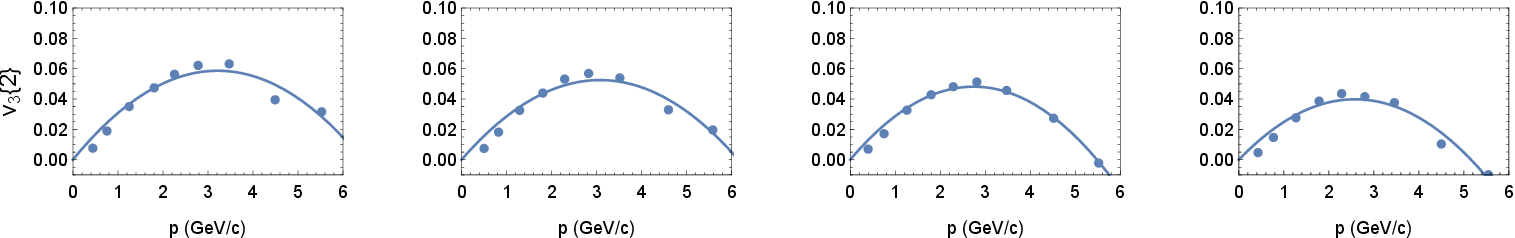}
    \caption{
Central values of $v_{2}\left \{ 2 \right \}$ (upper row), $v_{2}\left \{ 4 \right \}$ (middle row), and $v_{3}\left \{ 2 \right \}$ (lower row) as a function of transverse momentum $p$ in four $N_{ch}$ ranges (columns) in p-Pb collisions at $\sqrt{s_{NN}} = 5.02\ \mathrm{TeV}$ from the CMS experiment~\cite{cmsvn}. The curves represent polynomial fits to the experimental data.    
    }
    \label{vn}
\end{figure*}

    \subsection{The effect of $v_{2F}$ on $r_2$ and $r_3$}
\label{secv2F}
\renewcommand{\theequation}{C.\arabic{equation}}
\renewcommand{\thefigure}{C.\arabic{figure}} 
\renewcommand{\thetable}{C.\arabic{table}} 

\setcounter{equation}{0}  
\setcounter{figure}{0}     
\setcounter{table}{0}

\begin{figure*}[!htb]
    \centering
        \includegraphics[scale=0.7]{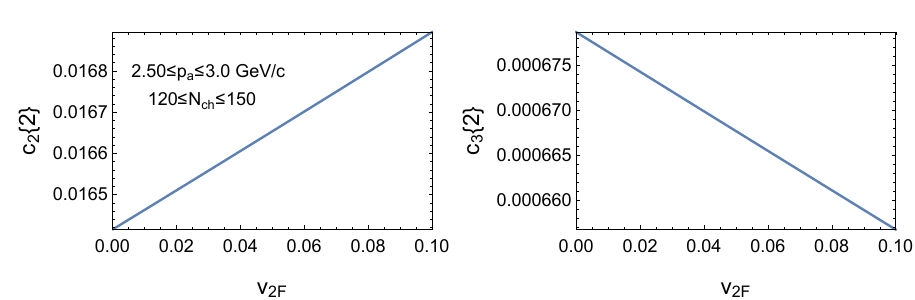}  
    \caption{$c_2\{2\}$ (left) and $c_3\{2\}$ (right) as a function of $v_{2F} $ for $2.5\le p_{a}\le 3.0 $ GeV/c and $120\le N_{ch}\le 150 $, where both $p_a$ and $N_{ch} $ are taken as their maximum values. }
    \label{c23v2F}
\end{figure*}

\begin{figure*}[!htb]
    \centering
        \includegraphics[scale=0.7]{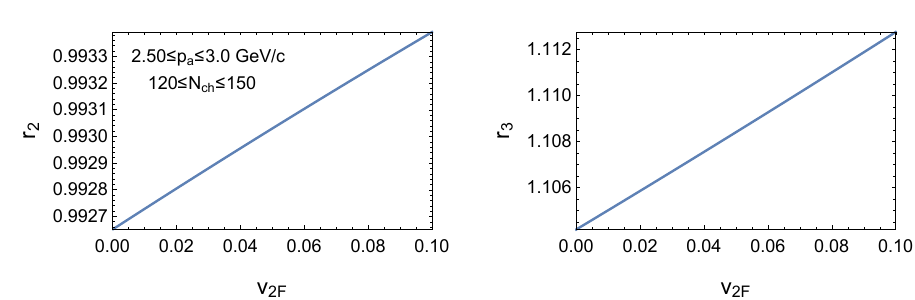}  
    \caption{$r_2$ (left) and $r_3$ (right) as a function of $v_{2F} $ for $2.5\le p_{a}\le 3.0 $ GeV/c and $120\le N_{ch}\le 150 $, where both $p_A$ and $N_{\text{ch}}$ are taken as their maximum values, and $p_A-p_B$ is set to 1 GeV/c.}
    \label{r23v2F}
\end{figure*}

In this part, we have already demonstrated that the results depend very weakly on $v_{2F}$, as shown in Figs.~\ref{c23v2F} and~\ref{r23v2F}. From Figs.~\ref{c23v2F} and~\ref{r23v2F}, it can be seen that the maximum effect induced by $v_{2F}$ is only on the order of one part in a thousand. This is because, for $c_2\{2\}$, the dominant contributions come from pure TMC and pure flow, while the terms related to $v_{2F}$ all appear in the interplay between them. For $c_3\{2\}$, besides pure TMC and pure flow, the dominant contributions also include a first-order interplay term that is independent of $v_{2F}$, whereas the terms related to $v_{2F}$ only appear in higher-order interplay terms.


    \subsection{$\cos\left [ n\Psi _{n} (p_{a} )-n\Psi _{n} (p_{b} ) \right ] $}
\label{secb}
\renewcommand{\theequation}{D.\arabic{equation}}
\renewcommand{\thefigure}{D.\arabic{figure}} 
\renewcommand{\thetable}{D.\arabic{table}} 

\setcounter{equation}{0}  
\setcounter{figure}{0}     
\setcounter{table}{0}

\begin{figure*}[!htb]
\centering
    \includegraphics[scale=1.0]{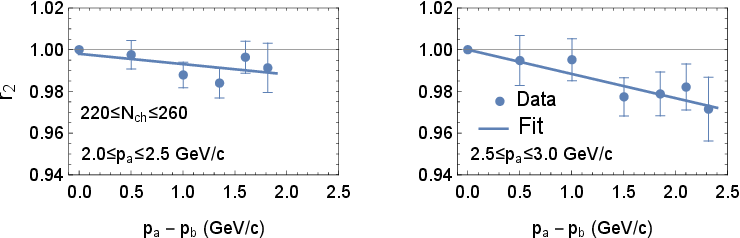}
    \caption{
Factorization ratio $r_2$ as a function of $p_{a}-p_{b}$ in two $p_a$ bins for high $N_{ch}$ range of $220\le N_{ch}\le 260 $ in p-Pb collisions at $\sqrt{s_{NN}} = 5.02\ \mathrm{TeV}$ from the CMS experiment~\cite{cmsdata}. The curves represent polynomial fits to the experimental data.
    }
    \label{cos}
\end{figure*}

Here we perform a polynomial fit to the experimental data of $r_2$ under high multiplicities. The specific fitting formulas are as follows:
\\
\\
$2.0\le p_{a} \le2.5 \: \text{GeV/c} $:
\begin{equation}
   \cos   \left [ 2\Psi _{2} (p_{a} )-2\Psi _{2} (p_{b} ) \right ]\approx r_{2} \approx 0.998-0.005\times \left ( p_{a}-p_{b}   \right ) , 
   \label{eqB1}
\end{equation}
$2.5\le p_{a} \le3.0 \: \text{GeV/c} $:
\begin{equation}
   \cos   \left [ 2\Psi _{2} (p_{a} )-2\Psi _{2} (p_{b} ) \right ]\approx r_{2} \approx 1-0.012\times \left ( p_{a}-p_{b}   \right ) .
   \label{eqB2}
\end{equation}
For $r_2$ at the remaining $p_a$ momentum bins and for $r_3$ at all $p_a$ momentum bins, the data are consistent with unity within experimental uncertainties. Therefore, for simplicity, we set these event-plane correlations to a value of 1 in our analysis. A comparison between the curves for Eqs.~\eqref{eqB1}, ~\eqref{eqB2}, and the experimental data is shown in Fig.~\ref{cos}. Here, we adopt the simplest possible fit that is consistent with the experimental uncertainties while reproducing the general trend of the data.

\subsection{$r_2$ and $r_3$ without considering $p_T$-dependent event plane fluctuations}
\label{secc}
\renewcommand{\theequation}{E.\arabic{equation}}
\renewcommand{\thefigure}{E.\arabic{figure}} 
\renewcommand{\thetable}{E.\arabic{table}} 

\setcounter{equation}{0}  
\setcounter{figure}{0}     
\setcounter{table}{0}

\begin{figure*}[!htb]
    \centering
        \includegraphics[scale=0.7]{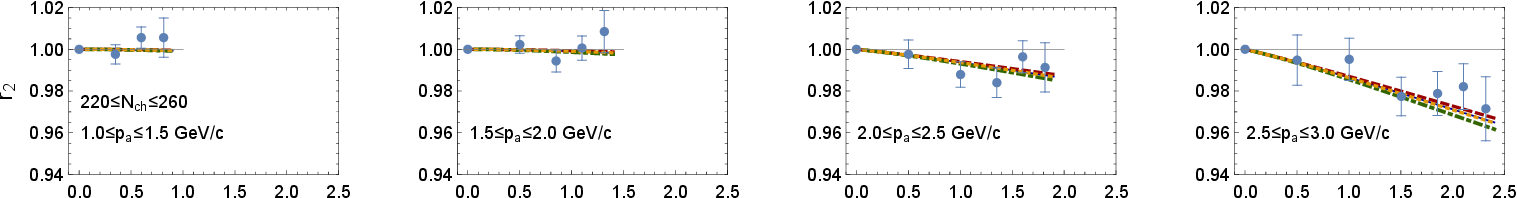}     \includegraphics[scale=0.7]{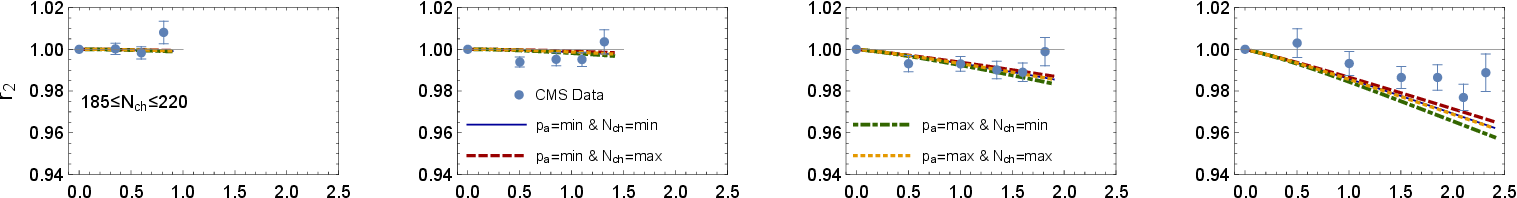}
    \includegraphics[scale=0.7]{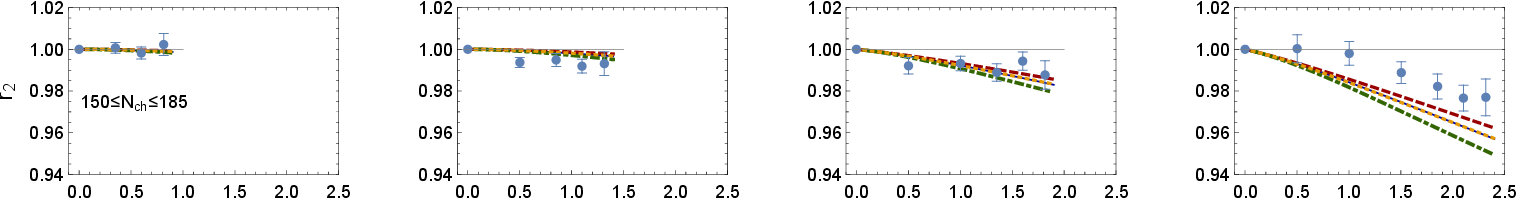}
\includegraphics[scale=0.7]{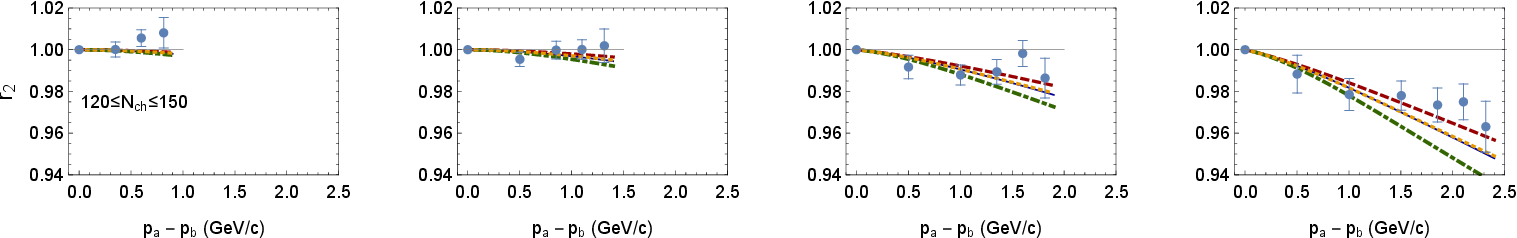}
    \caption{Same as Fig.~\ref{cos0r2}, but including $p_T$-dependent $\Psi_2$ fluctuations. }
    \label{r2}
\end{figure*}

\begin{figure*}[!htb]
    \centering
        \includegraphics[scale=0.7]{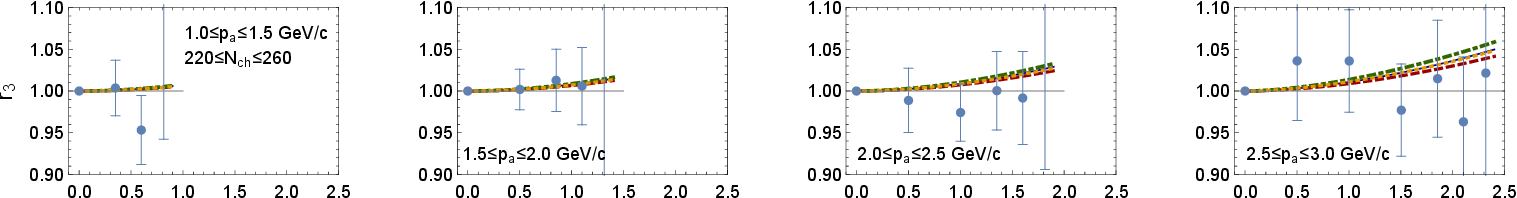}
    \includegraphics[scale=0.7]{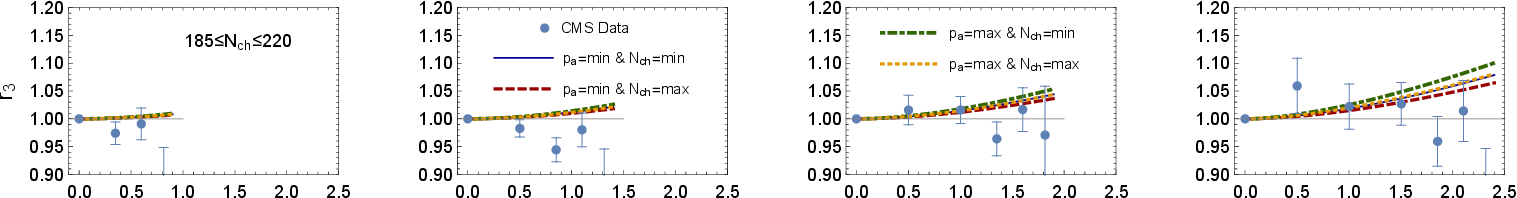}
\includegraphics[scale=0.7]{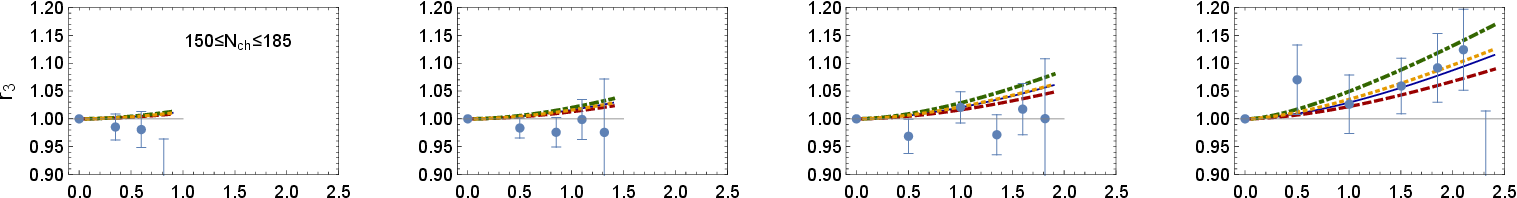}
   \includegraphics[scale=0.7]{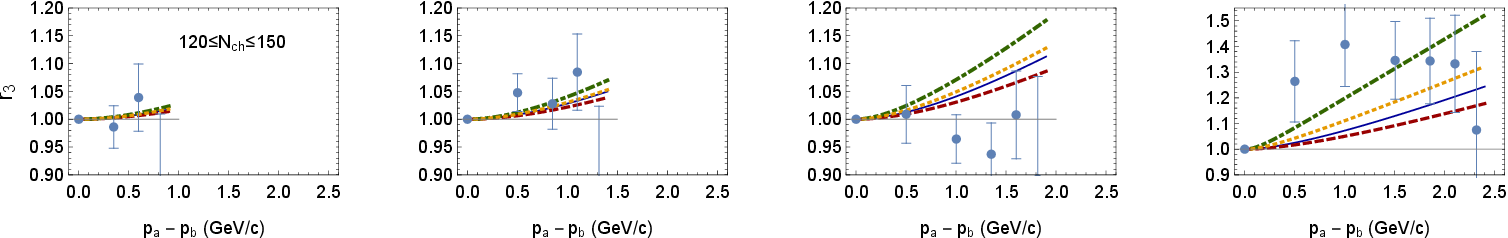}
    \caption{Same as Fig.~\ref{cos0r3}, but including $p_T$-dependent $\Psi_2$ fluctuations. }
    \label{r3}
\end{figure*}

Figures~\ref{r2} and~\ref{r3} show the results obtained by considering the $p_T$-dependent event plane fluctuations. Although this considers event-plane fluctuations, we find that both the quality of the fits and our key conclusions remain robust, compared to Figs.~\ref{cos0r2} and~\ref{cos0r3}.
Compared with Fig.~\ref{cos0r2}, the curves in Fig.~\ref{r2} are slightly suppressed. However, Fig.~\ref{r3} is virtually identical to Fig.~\ref{cos0r3}, indicating that $r_3$ is insensitive to fluctuations in the second-order event plane.

\subsection{Di-jet}
\label{secjet}
\renewcommand{\theequation}{F.\arabic{equation}}
\renewcommand{\thefigure}{F.\arabic{figure}} 
\renewcommand{\thetable}{F.\arabic{table}} 

\setcounter{equation}{0}  
\setcounter{figure}{0}     
\setcounter{table}{0}
In this part, we will calculate $r_n$ for the following three scenarios (with $\eta$ gap applied): ``flow+TMC'' (which is the result presented in Figs.~\ref{cos0r2} and~\ref{cos0r3}), ``flow+di-jet'', and ``flow+TMC+di-jet''.
\begin{widetext}
    \vspace{\baselineskip}
\noindent
\textbf{Flow + di-jet:} When both flow and di-jet contributions are present, the two-particle transverse momentum distribution is as follows.
\begin{equation}
 \begin{split}
        f_{2}(\vec{p}_{a},\vec{p}_{b}) =&
\left[
\frac{C}{2\pi} \left(1 + 2v_2(p_a)\cos 2\phi_1^a + 2v_3(p_a)\cos 3\phi_1^a\right)
+ J_1 \delta(\phi_1^a - \Psi_J) + J_2 \delta(\phi_1^a - \Psi_J - \pi)
\right] \\
&\times
\left[
\frac{C}{2\pi} \left(1 + 2v_2(p_b)\cos 2\phi_1^b + 2v_3(p_b)\cos 3\phi_1^b\right)
+ J_1 \delta(\phi_1^b - \Psi_J) + J_2 \delta(\phi_1^b - \Psi_J - \pi)
\right].
\end{split}
\label{f2tmcjet}
\end{equation}
Using
\begin{equation}
    c_{n}\left \{ 2 \right \}  \vert_{p_{a},p_{b}}  =\left \langle e^{in\left ( \phi _{1}^a-\phi _{2}^b\right ) } \right \rangle\big| _{p_{a},p_{b}} =
\frac{\int_{0}^{2\pi}e^{in\left ( \phi _{1}^a-\phi _{2}^b\right ) }
f_{2}(\vec{p}_{a},\vec{p}_{b})d\phi _{1}^a d\phi _{2}^b}
{\int_{0}^{2\pi}f_{2}(\vec{p}_{a},\vec{p}_{b})d\phi_{1}^a d\phi _{2}^b} ,
\end{equation}
where $f_{2}(\vec{p}_{a},\vec{p}_{b})$ takes the form of Eq.~\eqref{f2tmcjet}, we have
\begin{equation}
    \begin{split}
 c_{2}\left \{ 2 \right \}  \vert_{p_{a},p_{b}} &= \frac{C^2 v_2(p_a) v_2(p_b) + C(J_1 + J_2)\bigl[v_2(p_a) + v_2(p_b)\bigr]\cos  \left ( 2\Psi_J \right ) + (J_1 + J_2)^2}{(C + J_1 + J_2)^2}, \\
 c_{3}\left \{ 2 \right \}  \vert_{p_{a},p_{b}} &= \frac{C^2 v_3(p_a) v_3(p_b) + C(J_1 - J_2)\bigl[v_3(p_a) + v_3(p_b)\bigr]\cos  \left ( 3\Psi_J \right )  + (J_1 - J_2)^2}{(C + J_1 + J_2)^2}.
    \end{split}
    \label{cntmcjet}
\end{equation}
By removing the terms related to intra-jet correlations ($J_{1}^{2} +J_{2}^{2}$) in Eq.~\eqref{cntmcjet}, we obtain results analogous to those with an $\eta$ gap applied. Furthermore, the jet orientation $\Psi_J$ in Eq.~\eqref{cntmcjet} is naturally assumed to be random, so the $\cos \left ( n\Psi_J \right )$ terms vanish after event averaging. The final results are as follows:
\begin{equation}
 c_{2}\left \{ 2 \right \}  \vert_{p_{a},p_{b}} = \frac{C^2 v_2(p_a) v_2(p_b) + 2J_1  J_2}
{C^2+2C\left ( J_1 + J_2 \right )+2J_1  J_2 } , \hspace{0.5cm}
 c_{3}\left \{ 2 \right \}  \vert_{p_{a},p_{b}} = \frac{C^2 v_3(p_a) v_3(p_b) - 2J_1  J_2}
{C^2+2C\left ( J_1 + J_2 \right )+2J_1  J_2 }.
    \label{cntmcjetf}
\end{equation}
Based on Eq.~\eqref{cntmcjetf}, $r_n$ can be obtained:
\begin{equation}
r_{2} =\frac{C^2 v_2(p_a) v_2(p_b) + 2J_1  J_2}
{\sqrt{\left ( C^2 v_2(p_a)^{2}  + 2J_1  J_2 \right )\times \left ( C^2 v_2(p_b)^{2}  + 2J_1  J_2 \right )}  }  ,\hspace{0.3cm}
r_{3} =\frac{C^2 v_3(p_a) v_3(p_b) - 2J_1  J_2}
{\sqrt{\left ( C^2 v_3(p_a)^{2}  - 2J_1  J_2 \right )\times \left ( C^2 v_3(p_b)^{2}  -2J_1  J_2 \right ) }} .
\end{equation}

\vspace{\baselineskip}
\noindent
\textbf{Flow + TMC + di-jet}: When flow, TMC, and di-jet contributions are all present, the two-particle transverse momentum distribution is given by:
\begin{equation}
 \begin{split}
        f_{2}(\vec{p}_{a},\vec{p}_{b}) =&
\left[
\frac{C}{2\pi} \left(1 + 2v_2(p_a)\cos 2\phi_1^a + 2v_3(p_a)\cos 3\phi_1^a\right)
+ J_1 \delta(\phi_1^a - \Psi_J) + J_2 \delta(\phi_1^a - \Psi_J - \pi)
\right] \\
&\times
\left[
\frac{C}{2\pi} \left(1 + 2v_2(p_b)\cos 2\phi_1^b + 2v_3(p_b)\cos 3\phi_1^b\right)
+ J_1 \delta(\phi_2^b - \Psi_J) + J_2 \delta(\phi_2^b - \Psi_J - \pi)
\right] \\
& \times \frac{N}{N-2} \text{exp}\left ( -\frac{ p_{a}^{2} +p_{b}^{2} +2p_{a}p_{b}\cos \left ( \phi _1^a-\phi _2^b  \right ) }
{(N-2)\left \langle p^{2}  \right \rangle_{F} }  \right ).
\end{split}
\label{f2flowtmcjet}
\end{equation}
Similarly, we can obtain $c_n\{2\}$ in this case (after removing intra-jet correlations):
\begin{equation}
    \begin{split}
  c_{2}\{2\}\big|_{p_{a},p_{b}} =& \frac{1}{Y}\bigg( 2 J_1 J_2 + C^2 v_{2}(p_a) v_{2}(p_b) - 2 J_1 J_2 (p_a - p_b)^2 x - C^2 (p_a^2 + p_b^2) v_{2}(p_a) v_{2}(p_b) x \\
  &+ \frac{1}{2} \Bigl(2 J_1 J_2 (p_a - p_b)^4 x^2 
  + 2 C J_1 p_a^2 p_b^2 x^2 + 2 C J_2 p_a^2 p_b^2 x^2 \\
  &+ C^2 \bigl(p_a^4 v_{2}(p_a) v_{2}(p_b) + p_b^4 v_{2}(p_a) v_{2}(p_b) + p_a^2 p_b^2 (1 + 4 v_{2}(p_a) v_{2}(p_b))\bigr) x^2\Bigr) \bigg),
    \end{split}
    \label{c2all}
\end{equation}
\begin{equation}
\begin{split}
    c_{3}\{2\}\big|_{p_{a},p_{b}} =&\frac{1}{Y}\bigg(-2 J_1 J_2 + C^2 v_{3}(p_a) v_{3}(p_b) + 2 J_1 J_2 (p_a - p_b)^2 x \\
   & - C^2 \bigl(p_a p_b v_{2}(p_a) v_{2}(p_b) + p_a^2 v_{3}(p_a) v_{3}(p_b) + p_b^2 v_{3}(p_a) v_{3}(p_b)\bigr) x \\
    &+ \frac{1}{2} \Bigl(-2 J_1 J_2 (p_a - p_b)^4 x^2 + C^2 \bigl(2 p_a p_b (p_a^2 + p_b^2) v_{2}(p_a) v_{2}(p_b) + (p_a^4 + 4 p_a^2 p_b^2 + p_b^4) v_{3}(p_a) v_{3}(p_b)\bigr) x^2\Bigr) \\
    &+ \frac{1}{6}  \Bigl(2 J_1 J_2 (p_a - p_b)^6 x^3 - 2 C J_1 p_a^3 p_b^3 x^3 - 2 C J_2 p_a^3 p_b^3 x^3 - C^2 \bigl(3 p_a^5 p_b v_{2}(p_a) v_{2}(p_b) + 3 p_a p_b^5 v_{2}(p_a) v_{2}(p_b) \\
    &+ p_a^3 p_b^3 (1 + 9 v_{2}(p_a) v_{2}(p_b)) + p_a^6 v_{3}(p_a) v_{3}(p_b) + 9 p_a^4 p_b^2 v_{3}(p_a) v_{3}(p_b) + 9 p_a^2 p_b^4 v_{3}(p_a) v_{3}(p_b) + p_b^6 v_{3}(p_a) v_{3}(p_b)\bigr) x^3\Bigr)\bigg),
\end{split}
\label{c3all}
\end{equation}
where
\begin{equation}
    x=\frac{1}{(N-2)\left \langle p^{2}  \right \rangle_{F}} ,\hspace{0.5cm} Y=C^2+2C\left ( J_1 + J_2 \right )+2J_1  J_2 .
\end{equation}
\end{widetext}
Due to the complex form of Eq.~\eqref{f2flowtmcjet}, we perform the calculation using the Taylor expansion of the exponential function, as done in the manuscript. Furthermore, according to Eqs.~\eqref{A} and~\eqref{B} in Sec. \ref{TMC calculations}, the result for $c_2\{2\}$ is obtained by expanding the exponential term to the second order, while the result for $c_3\{2\}$ is obtained by expanding the exponential term to the third order. Hence, Eqs.~\eqref{c2all} and~\eqref{c3all} are obtained by expanding the exponential term in Eq.~\eqref{f2flowtmcjet} to the second and third orders, respectively.

\begin{figure*}[!htb]
 \centering
    \includegraphics[scale=0.6]{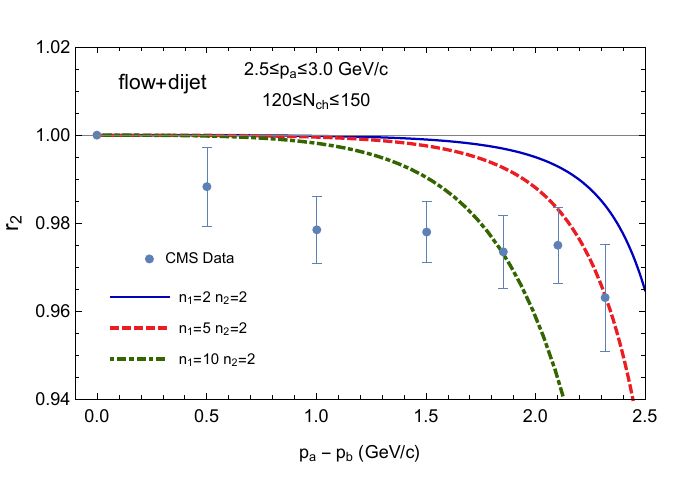}
    \includegraphics[scale=0.6]{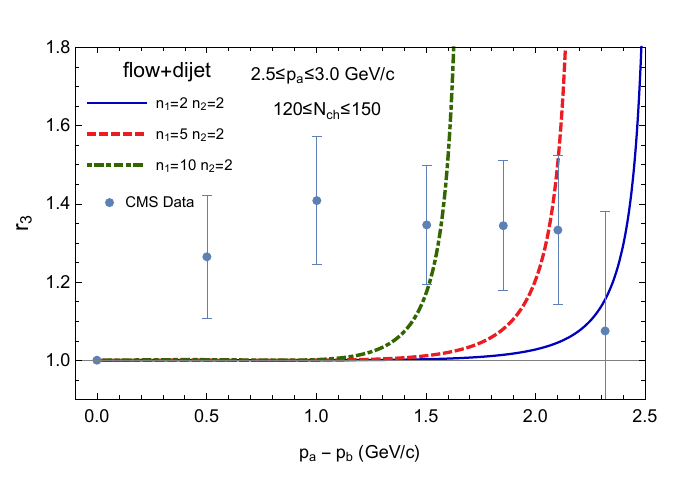}
    \includegraphics[scale=0.6]{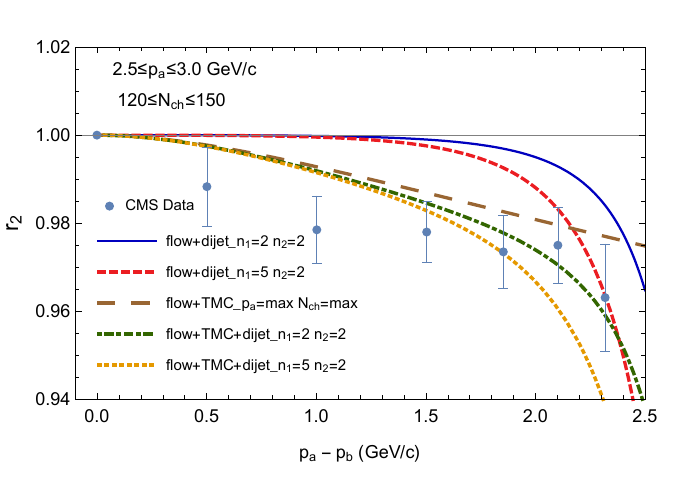}
    \includegraphics[scale=0.6]{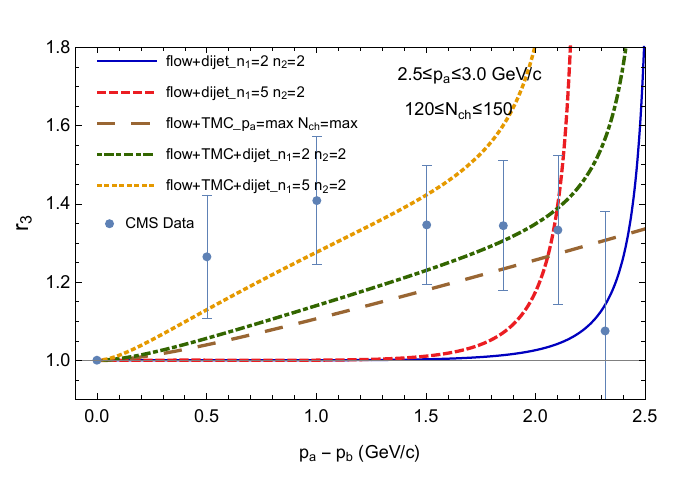}
    \caption{Factorization ratio $r_2$ (left column) and $r_3$ (right column) as a function of $p_{a}-p_{b}$ for $2.5\le p_{a}\le 3.0 $ GeV/c and $120\le N_{ch}\le 150 $. Curves: calculations from flow+di-jet, flow+TMC, and flow+TMC+di-jet. Points: CMS p-Pb data at 5.02 TeV with statistical errors (systematic uncertainties negligible)~\cite{cmsdata}.}
    \label{rnall}
\end{figure*}

With all of that, Fig.~\ref{rnall} is obtained based on Eq.~\eqref{cntmcjetf},~\eqref{c2all},~\eqref{c3all}, and the result presented in Figs.~\ref{cos0r2} and~\ref{cos0r3}).
It can be seen from Fig.~\ref{rnall} that when the non-flow contribution only comes from di-jet, it can indeed lead to $r_2<1$ and $r_3>1$. But significant deviations only appear at large transverse momentum differences, while at small transverse momentum differences, the results are far from the experimental data points.
We find that the TMC and dijet contributions could remain coherently present in the observed $r_2$ and $r_3$. However, they are not degenerate, as their relative importance changes with the transverse momentum difference. Unless the TMC contribution is included, the CMS data for $r_2$ and $r_3$ fail to be well described theoretically, especially at small transverse momentum differences. 

\end{document}